\documentclass[12pt,epsfig,color]{article}
\usepackage{amsmath}
\usepackage{amsfonts}
\usepackage{amsmath}
\usepackage{slashed}
\usepackage{amsxtra}
\usepackage{amstext}
\usepackage{amssymb}
\usepackage{color}
\usepackage{float}
\usepackage{graphicx}
\usepackage{subcaption}
\numberwithin{equation}{section}
\usepackage[dvips]{epsfig}
\usepackage{appendix}
\newcommand{\be}{\begin{equation}}
\newcommand{\ee}{\end{equation}}
\newcommand{\beq}{\begin{equation}}
\newcommand{\eeq}{\end{equation}}
\newcommand{\ba}{\begin{eqnarray}}
\newcommand{\ea}{\end{eqnarray}}
\newcommand{\bea}{\begin{eqnarray}}
\newcommand{\eea}{\end{eqnarray}}
\newcommand{\nn}{\nonumber}
\newcommand{\al}{\alpha}
\newcommand{\vep}{\varepsilon}

\newcommand{\der}{\partial}

\newcommand{\m}{\mu}
\newcommand{\n}{\nu}

\begin{document}
\baselineskip=15.5pt \pagestyle{plain} \setcounter{page}{1}
%
\begin{titlepage}

\vskip 0.8cm

\begin{center}
%
%
%
%
%

{\Large \bf Scope and limitations of a string theory dual description of the proton structure}

\vskip 1.cm

{\large {{\bf David Jorrin}{\footnote{\tt jorrin@fisica.unlp.edu.ar}} {\bf and  Martin
Schvellinger}{\footnote{\tt martin@fisica.unlp.edu.ar}}}}

\vskip 1.cm

{\it Instituto de F\'{\i}sica La Plata-UNLP-CONICET. \\
Boulevard 113 e 63 y 64, (1900) La Plata, Buenos Aires, Argentina \\
and \\
Departamento de F\'{\i}sica, Facultad de Ciencias Exactas,
Universidad Nacional de La Plata. \\
Calle 49 y 115, C.C. 67, (1900) La Plata, Buenos Aires, Argentina.}

\vspace{1.cm}

{\bf Abstract}

\end{center}

\vspace{1.cm}

Symmetric and antisymmetric structure functions from  electromagnetic deep inelastic scattering of charged leptons off  spin-1/2 hadrons are investigated in the framework of a top-down holographic dual description. We consider the BPST Pomeron, type IIB superstring theory scattering amplitudes, and type IIB supergravity on AdS$_5 \times S^5$. In all cases it is used the hard-wall prescription. Different kinematic regions of the Bjorken variable $x$, as well as the squared momentum of the virtual photon $Q^2$, are studied in detail for $F_2^P$ and $g_1^P$ structure functions of the proton. Also, the virtual Compton scattering asymmetry of the proton $A_1^P$ is investigated. Comparison with data from several experimental collaborations is presented. In addition, the holographic Pomeron leads to predictions for the mentioned observables for very small $x$ values. In particular, we present predictions for $g_1^P$ at $Q^2$ around 10 GeV$^2$, for data expected to be measured in a future electron-ion collider. Limitations of this holographic dual approach are discussed.

\noindent

\end{titlepage}

\newpage

{\small \tableofcontents}

\newpage


%
\section{Introduction}\label{S-1}
%

Deep inelastic scattering (DIS) of leptons off hadrons is one of the most important experiments in the history of modern  high energy physics. First experiments of DIS of electrons off protons started at the two-mile accelerator at the Stanford Linear Accelerator Center in late 1967 \cite{Bloom:1969kc,Breidenbach:1969kd}. These experiments, together with the theoretical developments which encompassed those discoveries, led to a profound understanding of the structure of Nature within the domain of Quantum Chromodynamics. Very important further experimental and theoretical developments have produced an immense advance in the comprehension of the hadron structure\footnote{Several of these experimental collaboration are cited in section \ref{S-3} for the structure funtion $F_2$, in section \ref{S-4} for the antisymmetric function $g_1$, and in section \ref{S-5} for the virtual Compton scattering asymmetry of the proton.}. The next step towards the understanding of the hadron structure will be the experimental program at the Electron-Ion Collider (EIC). It will lead to the possibility of exploring very small values of the Bjorken parameter, $x$, and simultaneously a wide range of the squared virtual-photon momentum, $Q^2$. In this parametric region, the physics of the nucleon and nuclei structure is dominated by the gluons. It is also expected that the EIC will provide unprecedented access to the spatial and spin structure of the proton, neutron and light ions \cite{AbdulKhalek:2021gbh}. These are strong motivations to develop new models as well as to explore their ability to predict the behaviour of the hadron structure functions in this kinematic domain.

Our present work focuses on two very interesting aspects. On the one hand, it analyses the comparison of models derived from string theory, in terms of the gauge/string theory duality, with data for symmetric and antisymmetric structure functions from several experimental collaborations within the already explored kinematical ranges. We will see how using a very few parameters, many experimental data are fitted very well for small and moderately small values of the Bjorken parameter. On the other hand, the formulas used to fit data are also valid for a kinematic regime, where there are no experimental data yet (i.e. for very small $x$ and $Q^2$ around 10 GeV$^2$ for the antisymmetric structure function $g_1^P$). Thus, this also gives predictions for experimental data expected to be measured at the EIC. These are compelling reasons for the development of the work we present in this article.

~

For polarized charged leptons and polarized hadrons the DIS differential cross section corresponding to a final polarized lepton in the solid angle $d \Omega$ and in the final energy range $(E', E'+dE')$, is given by \cite{Anselmino:1994gn}
\begin{equation}
\frac{d^2 \sigma}{d \Omega \ dE'} =  \frac{\alpha_{em}^2}{2 M q^4} \ \frac{E'}{E} \ l_{\mu\nu} \ W^{\mu\nu} \, .
\end{equation}
This is in the laboratory frame where the hadron carries four-momentum $P_\mu = (M, 0)$, while the incoming and outgoing lepton four-momenta are $k_\mu = (E, \vec{k})$ and $k'_\mu = (E', \vec{k}')$, respectively. $M$ denotes the nucleon mass and $\alpha_{em}$ is the fine structure constant. This expression assumes the exchange of a single virtual photon between the incoming lepton and the hadron. The differential cross section is defined in terms of the so-called leptonic tensor $l_{\mu\nu}$ and the hadronic tensor $W^{\mu\nu}$. The virtual photon which probes the hadron structure carries four-momentum $q_\mu=k_\mu-k'_\mu$. The four-dimensional Minkowski metric is defined mostly plus: $\eta_{\mu\nu}=\text{diag}(-1,1,1,1)$. There is also a spin four-vector corresponding to the incoming baryon, $S_\mu$. In addition, the Bjorken variable is defined as
\begin{equation}
x=-\frac{q^2}{2P\cdot q}=\frac{Q^2}{2P\cdot q} \, ,
\end{equation}
where $0 \leq x \leq 1$ corresponds to its physical range and $Q^2=-q^2$. In the DIS limit $Q^2$ becomes very large, while $x$ is kept fixed. For a spin-$1/2$ baryon one may write the following decomposition for the hadronic tensor \cite{Anselmino:1994gn,Lampe:1998eu}
\be
W_{\mu\nu}=W^{\text{(S)}}_{\mu\nu}(q,P)+i \,
W^{\text{(A)}}_{\mu\nu}(q,P,S) \, , 
\ee
where the symmetric part is
\bea
W^{\text{(S)}}_{\mu\nu}&=& \left(\eta_{\mu\nu}-\frac{q_\mu
q_\nu}{q^2}\right) \left[F_1(x,q^2)+\frac{1}{2} \frac{S\cdot
q}{P\cdot q}g_5(x,q^2)\right] \, , \nn \\
&& -\frac{1}{P\cdot q} \left(P_\mu - \frac{P\cdot q}{q^2} q_\mu
\right) \left(P_\nu - \frac{P\cdot q}{q^2} q_\nu \right)
\left[F_2(x,q^2)+\frac{S\cdot q}{P\cdot q}g_4(x,q^2)\right] \nn\\
&&
-\frac{1}{2 P\cdot q}\left[\left(P_\mu - \frac{P\cdot q}{q^2} q_\mu
\right) \left(S_\nu - \frac{S\cdot q}{P\cdot q} P_\nu \right)+
\left(P_\nu - \frac{P\cdot q}{q^2} q_\nu \right)\left(S_\mu -
\frac{S\cdot q}{P\cdot q} P_\mu \right)\right] \nn \\
&& g_3(x,q^2) \, , \nn \\
\label{hadronic-tensor-WS}
\eea
and the antisymmetric part is given by
\bea
W^{\text{(A)}}_{\mu\nu}&=&-\frac{\vep_{\mu\nu\rho\sigma}q^\rho}{P\cdot q} \left\{S^\sigma g_1(x,q^2) + \left[S^\sigma-\frac{S\cdot q}{P\cdot q}P^\sigma\right]g_2(x,q^2)\right\}
-\frac{\vep_{\mu\nu\rho\sigma}q^\rho P^\sigma}{2P \cdot q} F_3(x,q^2) \, . \nn \\
&&
\label{hadronic-tensor-WA}
\eea
Notice that in QCD for the electromagnetic DIS the functions $g_3$, $g_4$, $g_5$ and $F_3$ do not appear. On the other hand, for ${\cal {N}}=4$ supersymmetric Yang-Mills theory, with a certain kind of IR deformation, $F_3$ is non-zero \cite{Hatta:2009ra,Kovensky:2017oqs,Kovensky:2018xxa}. This IR deformation is such that there are massless Nambu-Goldstone modes emerging from the spontaneous breaking of the $R$-symmetry \cite{Hatta:2009ra}.

The optical theorem based on the unitarity of the $S$-matrix relates the forward Compton scattering amplitude to the DIS cross section. Thus, there are the relations
\begin{equation}
W_{\m\n}^{({\text{S}})} = 2 \pi \
{\text{Im}}\left[T_{\m\n}^{({\text{S}})}\right]
\,\,\,\,\, {\text{and}} \,\,\,\,\,\   W_{\m\n}^{({\text{A}})} = 2 \pi \ {\text{Im}}\left[T_{\m\n}^{({\text{A}})}\right] \, ,
\end{equation}
where $T^{\m\n}$ is given by the time-ordered expectation value of the product of two electromagnetic currents inside the hadron \footnote{We also use the variable $x$ to represent the four-dimensional Minskowski spacetime coordinates $x\equiv x_\mu = (x_0, x_1. x_2, x_3)$.}
\begin{equation}
T_{\m\n}\equiv i \int d^{4}x \ e^{i q\cdot x} \langle P|
{{\hat{\text{T}}}} \{J_\mu(x) J_\nu(0)\} |P \rangle \, .
\end{equation}

In addition, for longitudinally polarized hadrons, the longitudinal spin-spin asymmetry $A_\parallel$ for lepton+proton$\rightarrow$lepton+$X$ can also be measured \cite{Anselmino:1994gn}.  This is constructed from differential scattering cross sections of electrons with parallel ($\rightarrow$) or anti-parallel ($\leftarrow$) spin aligned with respect to the direction of motion. Let us take it along the $x_3$ coordinate. On the other hand, protons can be polarized parallel ($\Rightarrow$) or anti-parallel ($\Leftarrow$) with respect to the direction of motion of the lepton beam\footnote{Also the hadrons can be perpendicularly polarized, both up ($\Uparrow$) or down ($\Downarrow$). We shall focus only on the longitudinally polarized case of both leptons and hadrons.}. The longitudinal spin-spin asymmetry is defined as
\begin{equation}
A_{\parallel} = \frac{d\sigma^{\rightarrow}_{\Leftarrow}-d\sigma^{\rightarrow}_{\Rightarrow}}{d\sigma^{\rightarrow}_{\Leftarrow}+d\sigma^{\rightarrow}_{\Rightarrow}} \, . 
\end{equation}
In order to simplify some equations the notation has been abbreviated by defining the differential cross sections as $d\sigma = d^2\sigma/(d\Omega \ dE')$ with arrows indicating the corresponding polarization states. The longitudinal spin-spin asymmetry can be written from the virtual Compton scattering asymmetries $A_1$ and $A_2$ as
\begin{equation}
A_{\parallel} = D (A_1 + \eta A_2) \, , \label{A_parallel}
\end{equation}
where
\begin{equation}
A_1 = \frac{g_1-(4 M^2 x^2/Q^2) g_2}{F_1} \, ,
\end{equation}
which is a function that we study in section \ref{S-5} and compare with experimental data for the proton $A_1^P$, and also
\begin{equation}
A_2 =\frac{2 M x}{\sqrt{Q^2}} \ \frac{g_1+g_2}{F_1} \, ,
\end{equation}
where $F_1$ is a symmetric structure function in equation (\ref{hadronic-tensor-WS}), while $g_1$ and $g_2$ are antisymmetric structure functions in (\ref{hadronic-tensor-WA}). In addition, $D$ and $\eta$ in equation (\ref{A_parallel}) are given by
\begin{equation}
D = \frac{E - \epsilon E'}{E(1+\epsilon R)} \, ,
\end{equation}
and
\begin{equation}
\eta = \frac{\epsilon \sqrt{Q^2}}{E - \epsilon E'} \, ,
\end{equation}
while $\epsilon$ in the two previous equations is defined as
\begin{equation}
\epsilon = \frac{1}{1+2 \left(1+\frac{\nu^2}{Q^2}\right) \tan^2(\theta/2)} \, ,
\end{equation}
with $\nu = E-E'$, while 
\begin{equation}
R = \frac{F_2}{2 x F_1} \left(1+\frac{4 M^2 x^2}{Q^2}\right)-1  \, ,
\end{equation}
has been defined as the ratio of the longitudinal to transverse cross sections.

Since in the DIS limit $\eta$ and $A_2$ are very small, from equation (\ref{A_parallel}) we can write
\begin{equation}
A_{\parallel} \approx D \ A_1 \, ,
\end{equation}
and within the same approximation $R$ becomes
\begin{equation}
R \approx \frac{F_2-2 x F_1}{2 x F_1} \, .
\end{equation}
Finally, one obtains
\begin{equation}
A_1 \approx 2 x \ (1+R) \ \frac{g_1}{F_2} \, ,
\end{equation}
which is to be compared with $A_1^P$ in section \ref{S-5}.

~

In order to calculate the mentioned relevant quantities related to observables, the problem is how to calculate the tensor $T_{\m\n}$, taking into account the non-perturba\-ti\-ve effects due to QCD soft-processes. There are several approaches for different parametric regions in term of $Q^2$ and the Bjorken parameter (for a review see for instance the books \cite{Devenish:2004pb,Forshaw:1997dc,Donnachie:2002en}). Specially important is the DGLAP formulation where the splitting functions, written in terms of the gluon Bremsstrahlung by quarks and the quark anti-quark pair production from a gluon, play a fundamental role \cite{Dokshitzer:1977sg,Gribov:1972ri,Altarelli:1977zs}. 
Particularly, for scattering at small angles and high energies, the description involves a soft-Pomeron Regge pole corresponding to a glueball and a hard BFKL Pomeron which emerges from the leading order QCD calculations at weak coupling \cite{Lipatov:1976zz,Kuraev:1977fs,Balitsky:1978ic}. In \cite{Brower:2006ea} it has been constructed a unified description of the soft and hard Pomerons. The resulting object is known as the BPST Pomeron, which for negative values of the $t$-channel Mandelstam variable leads to results similar to those obtained from the BFKL Pomeron. On the other hand, for positive $t$-values it gives the expected Regge behaviour. The BPST Pomeron is based on the gauge/string theory duality. In the context of DIS, this duality has been developed by Polchinski and Strassler in the pioneering article \cite{Polchinski:2002jw}. They firstly considered the supergravity regime, where the $s$-channel dominates both for glueballs and spin-1/2 fermions, and then studied the small-$x$ region where superstring theory scattering amplitudes provide the leading contribution in the large-$N_c$ limit of the dual gauge theory for DIS of charged leptons from glueballs. The calculation of the full hadronic tensor for a spin-1/2 hadron from type IIB superstring theory scattering amplitudes has been done in \cite{Kovensky:2018xxa}.

Also, hard scattering in the gauge/string theory duality framework was previously considered by these authors in \cite{Polchinski:2001tt}, obtaining a crucial result, namely: fundamental strings propagating in certain curved spaces lead to the correct power-law behaviour for high-energy scattering amplitudes of hadrons. In that particular case they considered type IIB closed strings propagating in AdS$_5 \times S^5$ (with  a sharp IR cut-off leading to a confinement scale in the dual gauge theory) representing the hard scattering of $2 \to m$ glueballs. The warp factor of the curved AdS space-time leads to the power-law behaviour for the scattering amplitude, which is totally different in comparison with the typical soft (exponentially decaying) behaviour obtained from propagation of strings in Minkowski space. The warp factor also provides a mechanism to understand the size of hadrons from a dual string theoretical perspective \cite{Polchinski:2001ju}, which is deeply related to the developments presented in \cite{Polchinski:2002jw}.

The calculation of the $F_2$ structure function from the BPST Pomeron has been originally done in \cite{Brower:2010wf}, and their results include the conformal case (where there is no IR cut-off), the hard-wall BPST Pomeron, and also the corrections coming from the eikonal approximation. Their main result was to show how good is the description of the small-$x$ range of DIS of data from HERA \cite{H1:2009pze} in terms of the exchange of a single BPST Pomeron. Considering the hard-wall BPST Pomeron they found that for a combined H1-ZEUS data set which originally contained 249 points, after excluding ``ouliers" by using a sieving method (with a $\Delta\chi^2_{\text{max}}=4$), the fit turns out to be quite good. They found a $\chi^2_{\text{d.o.f.}}$ per degree of freedom equal to 1.07 for the range 0.1 GeV$^2< Q^2 \leq $ 400 GeV$^2$ and the Bjorken parameter smaller than 0.01. For comparison with the new results that we have obtained in our present work, we display the mentioned fit of reference \cite{Brower:2010wf} in the second line of our table 1. One should notice that using the BPST Pomeron there are only four free parameters to fit all data in the mentioned kinematical ranges. Using the parameters of the $F_2$ fit obtained in \cite{Brower:2010wf}, one of the remarkable results of reference \cite{Kovensky:2018xxa} was to derive the contribution from the exchange of a holographic Pomeron to the calculation of the antisymmetric structure function $g_1$ and perform its comparison with data for the proton from the COMPASS collaboration \cite{COMPASS:2010wkz,COMPASS:2015mhb,COMPASS:2017hef}. In this case it appears only one additional parameter to fit 30 experimental points of reference \cite{COMPASS:2017hef}, obtaining $\chi^2_{\text{d.o.f.}}=1.074$, which is also a very good fit.

Thus, given the success of confronting the holographic calculations for both symmetric and antisymmetric structure functions with experimental data of the proton within 0.1 GeV$^2< Q^2 < $ 400 GeV$^2$ and $0 < x < 0.01$ ranges for $F_2$ \cite{Brower:2010wf} and  for $g_1$ \cite{Kovensky:2018xxa}, in this work we aim at exploring how well the fits behave by including more data from other experimental collaborations for both structure functions, and also considering a range of $x$ ten times larger than the $x$-range studied in these two previous papers. Specifically, for $F_2^P$ we consider an initial set of 305 points, in comparison with the initial data set used in \cite{Brower:2010wf} with 249 points of H1-ZEUS, for the same range $0 < x < 0.01$. Moreover, we then extend the range to $0.01 < x < 0.1$, which adds 204 points in this new range. This is explained in section \ref{S-3}. Then, in section \ref{S-4-2}, we also consider an extended parametric range for $g_1^P$, almost duplicating the number of data with respect to those included in \cite{Kovensky:2018xxa}, and still obtaining a very good comparison between theory and experiment.

We should emphasize that all situations in our analysis correspond to the full range of the virtual-photon momentum transfer 0.1 GeV$^2< Q^2 \leq $ 400 GeV$^2$, while the range of the Bjorken parameter is ten times larger than the one considered in \cite{Brower:2010wf} and \cite{Kovensky:2018xxa}, respectively. Thus, the number of experimental points in our present work is almost twice in comparison with the number of data considered in these references (going from 249 to around 500 points for $F_2^P$, and from 30 to more than 100 points for $g_1^P$, respectively). In addition, in section \ref{S-5} we further develop the holographic dual approach to investigate the virtual Compton scattering asymmetry of the proton, and compare it with experimental data, obtaining a good level of agreement. To our knowledge this is the first holographic dual study of $A_1^P$. We discuss different aspects of the limitations of this analysis. The main depart from experimental results occurs for the parametric region $0.1 < x < 1$, where valence quarks play a very important role, while the top-down holographic dual model we study does not describe fundamental quarks. We discuss these issues in the last section of the work.

%
\section{Spin-$1/2$ hadron structure from a top-down hologra\-phic dual approach}\label{S-2}
%

In this section we study predictions for the symmetric and antisymmetric structure functions of the proton from a holographic dual description based on type IIB superstring theory. Let us emphasize that there is not an specific holographic dual model of QCD, even in the large $N_c$ limit. This may seem an obstacle for describing real hadrons in terms of string theory dual models. However, there is a compelling reason to investigate the large $N_c$ limit of gauge theories like QCD from top-down models based on string theory: there are important properties of the hadron structure functions which are  ``universal", in the sense that they are independent of any specific holographic dual model\footnote{Universal properties from holographic mesons have been obtained for instance for the relations among different structure functions for scalar and vector mesons using very different holographic dual models \cite{Koile:2011aa,Koile:2013hba,Koile:2014vca,Koile:2015qsa}. In type IIA superstring theory these relations were calculated for the Sakai-Sugimoto model \cite{Sakai:2004cn} and for the D4D6 anti-D6-brane model \cite{Kruczenski:2003uq}. Also, in type IIB superstring theory in the case of the D3D7-brane model \cite{Kruczenski:2003be} it has been obtained the same relations. For spin-1/2 fermions in the supergravity limit see references \cite{Jorrin:2020cil,Jorrin:2020kzq}, while in the string theory and BPST-Pomeron regimes see the article \cite{Kovensky:2018xxa}.}. Thus, this ``universal" character should be reflected on the comparison of the holographic dual model with experimental data. Besides, in the case of strongly coupled quark-gluon plasma (QGPs) there is important agreement between the top-down holographic dual description based on type IIB superstring theory on an asymptotically AdS$_5$-Schwarzschild black hole times $S^5$ and lattice QCD calculations at finite temperature (for a review see \cite{Casalderrey-Solana:2011dxg}). For instance, this is for the case of mass transport properties such as the shear viscosity/entropy density ratio, both at extremely large 't Hooft coupling \cite{Kovtun:2004de}, and at finite coupling \cite{Buchel:2008ae}. Also, considering electric charge transport, such as for DC electrical conductivity at large coupling \cite{Caron-Huot:2006pee}, and in the strong coupling expansion \cite{Hassanain:2011fn}. We can also mention the photo-production rates \cite{Caron-Huot:2006pee}, including the strong coupling expansion \cite{Hassanain:2011ce,Hassanain:2012uj}, which enter the calculation of direct photon which can be compared with relativistic heavy-ion collisions experiments at the Relativistic Heavy Ion Collider and at the Large Hadron Collider. In addition, DIS of electrons off a QGPs has been studied in \cite{Hatta:2007cs}, while the strong coupling corrections have been derived in \cite{Hassanain:2009xw}. Another very important reason is that top-down models have only a few parameters inherited from the string theory side of the duality. These are related to the number of D3-branes, $N_c$, the fundamental string length squared, $\alpha'=l_s^2$, the normalization constants of the wave-functions of the bulk fields  or their corresponding Kaluza-Klein modes after dimensional reduction, and a certain cut-off to ensure IR confinement of the gauge theory. In this sense, top-down models are much more stringent than the bottom-up models like AdS/QCD.

We consider the holographic dual description in terms of the large $N_c$ limit of ${\cal {N}}=4$ SYM theory, with all the fields in the adjoint representation of $SU(N_c)$. Thus, this holographic dual model does not contain fermions in the fundamental representation. Therefore, one should expect to have a ``universal" description for the physics of DIS in the parametric region where the valence quarks of QCD are not relevant. This corresponds to low-$x$ values, where the dominant effects come from the gluon dynamics and the quark anti-quark sea in QCD. This is what we investigate in this work, trying to understand the results of the comparison with experiments. We also discuss certain aspects for large-$x$ values. In this case, however, rather focusing on the limitations of the model in that parametric region.

In this holographic dual model the baryon is represented by a 5-dimensional spin-1/2 Kaluza-Klein mode of the 10-dimensional dilatino ($\hat{\lambda}$) of type IIB supergravity, after dimensional reduction on $S^5$. We consider the low-lying Kaluza-Klein spin-1/2 fermion in AdS$_5$. By using the mapping of string/super\-gra\-vi\-ty states onto SYM operators, the corresponding ${\cal {N}}=4$ SYM theory operator is ${\cal {O}}_0^{(6)}(x) = C^{(6)} \, \text{Tr}(F_+ \lambda_{{\cal {N}}=4})(x)$. Its twist is $\tau=\Delta-s=3$, where $\Delta$ is the conformal dimension of an operator of spin $s$. This is a descendant operator of the ${\cal {N}}=4$ SYM theory obtained by the action of three supercharges on ${\cal {O}}_2^{I_2}(x) = C^{I_2}_{i_1 i_2} \, \text{Tr}(X_{i_1} X_{i_2})(x)$. The construction of these operators is done in terms of fields of the ${\cal {N}}=4$ SYM gauge supermultiplet, namely: 4 left Weyl fermions $\lambda_{{\cal {N}}=4}$; 6 real scalars $X_j$ with $j=1, \cdots , 6$; and $F_+$ representing the self-dual 2-form field strength.

Also, from the type IIB supergravity side we may consider the Kaluza-Klein modes for $k>0$ from the compactification on $S^5$. On the gauge field theory side, they correspond to local twist $\tau=k+3$ spin-1/2 fermionic operators ${\cal {O}}_k^{I_k, (6)}(x) = C^{I_k, (6)}_{i_1 \dots i_k} \, \text{Tr}(F_+ \lambda_{{\cal {N}}=4} X_{i_1} \dots X_{i_k})(x)$, where $I_k$ runs from 1 to the dimension of the irreducible representation of $SU(4)_R$. They belong to the $[1, k, 0]$ irreducible representation of the $R$-symmetry group. By increasing the number of scalar fields $k=0, 1, 2, 3, \cdots$ the dimension of the corresponding irreducible representation of $SU(4)_R$ group of these fermionic operators increases as 4, 20, 60, 140, $\cdots$.

A crucial point in order to calculate the DIS cross section is the relation to the imaginary part of the forward Compton scattering amplitude given by the optical theorem. The calculation of the FCS amplitude includes intermediate states. From the bulk gravitational theory point of view, the nature of the intermediate states depends on the particular kinematic region in which we are interested. There are three distinct parametric regions in terms of the relation between the 't Hooft coupling ($\lambda_{\text {'t Hooft}}$) and the Bjorken parameter $x$. Basically, different parametric regions depend upon the properties of the intermediate states in the FCS Feynman-Witten diagram. Firstly notice that in the large-$N_c$ limit only single hadron states contribute. This is represented in the holographic dual model as a single closed string. Thus, in terms of the 10-dimensional center-of-mass energy, ${\tilde{s}}$, there is the following relation:
\begin{equation}
{\tilde{s}} \lesssim \frac{(1-x)}{(4 \pi g_{\text{string}} N_c)^{1/2} \alpha' x} \, ,
\end{equation}
where $g_{\text{string}}$ is the string coupling and $\alpha'$ is the string constant. Notice that  $\lambda_{\text {'t Hooft}} \equiv g_{\text{string}} N_c$. In  the large $N_c$ limit and for $\lambda_{\text {'t Hooft}}^{-1/2} \ll x < 1$ only supergravity states can be excited. Therefore, the intermediate states in the SYM theory calculation just involve the fermionic single-trace operators  which we just have described, with certain selection rules worked out in \cite{Jorrin:2020cil,Jorrin:2020kzq}. In section \ref{S-2-1} we briefly review some results derived in these papers, which will be important in order to understand the limitations of the type IIB supergravity approach. At lower $x$ values, however, on the dual string theory side, massive type IIB string theory modes must be considered. Thus, the calculation is done in terms of four-closed strings scattering amplitude with two dilatinos and two gravitons. For the kinematic region of exponentially small $x$ the calculation can be performed by assuming the exchange of a single BPST Pomeron in the bulk theory. Notice that in the construction within the framework of the BPST Pomeron, hadrons are represented by their wave-functions which are approximated by Dirac-delta distributions as explained in \cite{Brower:2010wf}. The remarkable property of the BPST Pomeron is that it provides a unified framework containing both the soft Pomeron for positive $t$-values and the BFKL Pomeron for $t<0$.

Beyond the large $N_c$, one should calculate $1/N_c^2$ corrections. In the supergravity sector it implies the exchange of two Kaluza-Klein modes, which for the DIS process corresponds to two external states. Beyond supergravity, within string theory one should consider one-loop closed superstring scattering amplitudes. Furthermore, in the exponentially small-$x$ regime it should be necessary to study two BPST Pomerons exchange. In this parametric region eikonal methods are relevant for the description of DIS \cite{Cornalba:2006xm,Cornalba:2007zb,Brower:2007qh,Brower:2007xg,Cornalba:2008qf}\footnote{The symmetric structure $F_2$ has been studied within the eikonal approximation in \cite{Brower:2010wf}, in comparison with H1-ZEUS data. This non-linear approximation is related to the saturation effect and it is small for $Q^2 > 1$ GeV$^2$.}. Since we focus on the large $N_c$ limit, all the results discussed in this work correspond to tree-level calculations. General non-linear effects from the BPST-Pomeron kernel associated with several structure functions, as well as, the virtual Compton scattering asymmetry of the proton will be investigated elsewhere.

%
\subsection{A type IIB supergravity dual description of hadron structure functions}\label{S-2-1}
%

In order to obtain the hadronic tensor we have to calculate 
the expectation value of two electromagnetic currents inside the hadron. Notice that this can be expressed as the operator product expansion (OPE) of certain operators of the ${\cal {N}}=4$ SYM theory. At strong coupling and in the planar limit, this OPE is dominated by protected double-trace operators  \cite{Polchinski:2002jw}.

The metric of the AdS$_5 \times S^5$ space can be expressed as
\begin{equation}
ds^2= z^{-2} \, (dz^2 +  \eta_{\mu\nu} dx^\mu dx^\nu) + d\Omega_5^2 \, . \label{metric1}
\end{equation}
The radius of $S^5$ and AdS$_5$ is set to one. Indices $a, b,\cdots= 0,..., 4$ correspond to AdS$_5$ space. For its boundary space we use Greek letters $\mu,\nu,\cdots=0,..., 3$. In addition, five-sphere indices are denoted by Greek letters $\alpha, \beta, \cdots = 1,..., 5$. The radial coordinate $z \to 0$ in the UV. The hard-wall model contains an arbitrary IR cut-off at $z_0=1/\Lambda$ in order to induce color confinement in the dual gauge theory at the energy scale $\Lambda$.

The matrix element of two electromagnetic currents inside the hadron is obtained from the Gubser-Klebanov-Polyakov-Witten Ansatz. Thus, we have to evaluate the supergravity action on-shell, taking into account all possible intermediate states. 
The first step is to derive the effective five-dimensional supergravity action involving two dilatino fields and a massless vector field. This has been done in \cite{Jorrin:2020cil} from the covariant type IIB supergravity equations of motion. The relevant part of the action can be written as
\begin{eqnarray}
S_{int}&=& K \int dz \ d^4x \sqrt{-g_{AdS_5}} \times \nonumber \\
&& \ \left( i \frac{{\cal{Q}}}{3}
\bar{\lambda}^-_{k} \gamma^a B^1_a\lambda^-_{k} +i \ \frac{b^{-, -}_{1 k j}}{12} \bar{\lambda}^-_{j} F^{ab}  \Sigma_{ab} \lambda^-_k + i  \frac{b^{+, -}_{1 k j}}{12} \bar{\lambda}^+_{j} F^{ab} \Sigma_{ab} \lambda^-_k\right) \, , \label{five-dimensional-action} 
\end{eqnarray}
where $\lambda^{\pm}_{k}$ are the five-dimensional Kaluza-Klein modes of equation (\ref{spinor-expansion}) obtained from the dimensional reduction of the ten-dimensional dilatinos (\ref{10Ddilatinos}) on the five-sphere. Also, it has been defined the following constants involving angular integrals of spinor spherical harmonics on the five-sphere,
\begin{eqnarray}
b^{\pm,-}_{1 k j}&=&\left(1+ 2 \left(k\mp j+\frac{5}{2}\mp\frac{5}{2}\right) \right) \int d \Omega_5  (\Theta^{\pm}_{j})^{\dag} \tau_{\alpha}v^{\alpha}\Theta_{k} +4{\cal{Q}}  \int   d \Omega_5 (\Theta^{\pm}_{j})^{\dag}  \Theta^-_{k}  \, ,
\end{eqnarray}
where $\Theta^{\pm}_{k}$ are spinor spherical harmonics satisfying equation (\ref{thetachargeeigenstates}). $\tau_\alpha$ are the Gamma matrices and $v^\alpha$ are the Killing vectors on the five-sphere. The charge ${\cal {Q}}$ is given in equation (\ref{thetachargeeigenstates}).
The normalization constant $K$ in (\ref{five-dimensional-action}) is obtained from comparison with the type IIB supergravity action  \cite{DallAgata:1998ahf}. The massless vector field $B_a^1$ is a linear combination of off-diagonal fluctuations of the metric tensor and vector fluctuations of the Ramond-Ramond four-form potential,
\begin{equation}
B_a^1(x) \equiv A_a^1(x) - 16 \Phi_a^1(x) \, , \label{Bfield}
\end{equation}
being $A_a^1(x)$ the Kaluza-Klein modes obtained from the five-dimensional reduction of metric fluctuations,
\begin{equation}
h_{a \alpha} = \sum_{I_5} A_a^{I_5}(x) \, Y^{I_5}_\alpha(y) \, ,
\end{equation}
while $\Phi_a^1(x)$ are the Kaluza-Klein modes of the Ramond-Ramond four-form field fluctuations
\begin{equation}
a_{a \alpha\beta\gamma} = \sum_{I_5} \Phi_a^{I_5}(x)\, \epsilon_{\alpha\beta\gamma\delta\epsilon} \, \nabla^\delta Y^{I_5 \epsilon}(y) \, .
\end{equation}
Label $I_5$ stands for $(l_5,l_4,l_3,l_2,l_1)$ associated with the vector spherical harmonics on $S^5$, $Y^{I_5 \epsilon}(y)$. 
The masses of the vector fields $B_a^1$ are $M^2_{B, l}=l^2-1$ with $l \ge 1$. They transform in the {\bf 15}, {\bf 64}, {\bf 175}, $\cdots$ irreducible representations of $SU(4)$, for $l=1$, 2, 3, $\cdots$, respectively. Since in the gauge/gravity dual calculation we only consider the vector mode whose boundary value couples to the $U(1)_R$ $R$-symmetry current of the ${\cal {N}}=4$ SYM theory, we only need the corresponding massless vector modes, $B_a^1(x)$, satisfying the boundary condition
\begin{equation}
B^1_{\mu}(x, z\to0)= n_{\mu} \, e^{i q \cdot x} \, . \label{Bbc}
\end{equation}
while the solutions to the corresponding Maxwell-Einstein equations are
\begin{eqnarray}
B^1_{\mu}(x, z)=n_{\mu} \, e^{i q \cdot x}  \, q \, z \, K_1(q z), \ \ \ \ \ \  B^1_z(x, z)=i \, n \cdot q \, e^{i q \cdot x} \, z \,  K_{0}(q z) \, ,
\end{eqnarray}
where $K_{i}(q z)$ are the Bessel functions of second kind. The field strength is given by $F_{ab}=\nabla_a B_b^1-\nabla_b B_a^1$. Also we use $\Sigma_{ab}=\frac{1}{4}(\gamma_a \gamma_b-\gamma_b \gamma_a)$, where these Gamma matrices are defined on the AdS$_5$.

This dimensional reduction from first principles has been done in our previous work \cite{Jorrin:2020cil,Jorrin:2020kzq}. This procedure allows to calculate all the constants derived from explicitly solving angular integrals of the spinor spherical harmonics, thus obtaining selection rules for the Kaluza-Klein modes which take part in the interactions.

The ten-dimensional dilatino field can be written as 
\begin{equation}
    \hat{\lambda}(x, y)=\left({\begin{array}{c}
   0\\
   \lambda(x, y) \\
  \end{array} }\right) \, . \label{10Ddilatinos}
\end{equation}
Then, from the five-dimensional reduction one obtains the Kaluza-Klein modes
\begin{eqnarray}
\lambda(x, y)=\sum_k \left(\lambda^+_k(x) \Theta_k^+(y) + \lambda_k^{-}(x)
\Theta_k^-(y)\right) \, , \label{spinor-expansion}
\end{eqnarray}
where
\begin{eqnarray}
\tau^{\alpha}D_{\alpha} \Theta_k^{\pm}= \mp i
\left(k+\frac{5}{2} \right)\Theta_k^{\pm} \ \ \ \ \ \ \text{with} \ \
\  k \geq 0 \, . \label{thetakpm}
\end{eqnarray}
An important aspect is that the spinor spherical harmonics are charge eigentates, which can be seen from the following expression  
\begin{eqnarray}
\left( v^{\alpha}D_{\alpha} - \frac{1}{4} \tau^\alpha \tau^\gamma \nabla_\gamma v_\alpha \right) \Theta_k^{\pm}= - i {\cal {Q}} \, \Theta_k^{\pm} \, . \label{thetachargeeigenstates}
\end{eqnarray}
The five-dimensional masses of the Kaluza-Klein modes of the dilatino field $\lambda^{\pm}_k$ are $m^{\pm}_k$. In addition, $\pm$ indicate the two towers of masses associated with the  irreducible representations {\bf 4$^*$}, {\bf 20$^*$}, {\bf 60$^*$}, $\cdots$ ($-$), or {\bf 4}, {\bf 20}, {\bf 60}, $\cdots$ ($+$) of the $SO(6) \sim SU(4)$ isometry group. Here, we label coordinates $x$ on AdS$_5$ and $y$ on $S^5$.

The calculation of all the structure functions both symmetric and antisymmetric ones has been done in detail for fermionic operators of twist $\tau=3$ in \cite{Jorrin:2020cil} and for higher-twist operators of ${\cal {N}}=4$ SYM theory in \cite{Jorrin:2020kzq}. The most general expression is 
\begin{equation}
F_i = \beta^2_{m}F_i^m+\beta^2_P F^P_{i}+ \beta_m \beta_P F^{c}_{i}+\beta^2_{Pm} F^{P}_{i} +\beta^2_+ F^{P+}_i+\beta^2_-F^{P-}_i,
    \label{complete-structure-function}
\end{equation}
and similarly for $g_i$ structure functions. The $\beta$'s are coefficients obtained from angular integrals of spinor spherical harmonics $\Theta_k^{\pm}$. To understand the different contributions in equation (\ref{complete-structure-function}) let us recall that this is obtained from the optical theorem. Therefore, it has been calculated from the forward Compton scattering in the bulk of the AdS$_5$, which means that in each contribution there is the ``product" of two interaction vertices from the five-dimensional action (\ref{five-dimensional-action}) and a fermion internal propagator connecting them. $F_i^m$  represents the contribution from the minimal coupling on both vertices from the five-dimensional action. $F_i^P$ comes from the Pauli term in both vertices. $F_i^{Pm}$ includes one minimal coupling vertex and a Pauli vertex in the other interaction vertex, in the FCS Feynman-Witten diagram. In addition, $F^{P\pm}_i$ corresponds to the case where the internal fermion in the FCS diagram in AdS$_5$ has a quantum number $k\pm 1$, in comparison with the number of the external states (which is $k$ for both the incoming and outgoing states). The corresponding vertices are indicated in figure 1 of reference \cite{Jorrin:2020kzq}, while the full expressions for all the contributions to the structure functions are also detailed in that reference, and we shall not reproduce them here. In sections \ref{S-3-3} and \ref{S-4-3} we discuss the results of the fit for $F_2$ and $g_1$, in comparison to experimental data of the proton for this range of the Bjorken variable.

%
\subsection{DIS from type IIB superstring theory scattering amplitudes}\label{S-2-2}
%

Now, we focus on the parametric region $\exp{(-\lambda_{\text {'t Hooft}}^{1/2})} \ll x \ll \lambda_{\text {'t Hooft}}^{-1/2}$, which from the holographic dual perspective can be described by type IIB superstring theory scattering amplitudes. The calculation of all structure functions for spin-1/2 fermions has been done in reference \cite{Kovensky:2018xxa}. In this region, at strong coupling and large $N_c$, the holographic dual description of the DIS process, in principle, requires full closed string theory scattering amplitudes in AdS$_5 \times S^5$, which are unkown. Fortunately, the dominant $\tilde{t}$-channel contribution is well described by a local approximation where the closed string theory scattering amplitudes are calculated in ten-dimensional Minkowski space\footnote{This approximation was originally proposed in \cite{Polchinski:2002jw}, then developed for the holographic Pomeron in \cite{Brower:2006ea}, while in \cite{Kovensky:2017oqs} it was applied for the calculation of all the structure functions for the glueball. Furthermore, it was extended to spin-1/2 fermions in reference \cite{Kovensky:2018xxa}.}. The idea is that from the closed string theory scattering amplitude one can build up an effective Lagrangian from which it is possible to calculate the holographic dual FCS amplitude, and finally derive from it the structure functions. As in the previous subsection, we associate the ten-dimensional dilatino with the spin-1/2 fermionic operators of the gauge field theory, which in turn we will assume to represent the dual of the spin-1/2 hadrons. At this point one should recall that there is a difference in the holographic dual calculation of the symmetric and the antisymmetric structure functions for spin-1/2 fermions. While from this holographic dual viewpoint the symmetric structure functions can be derived from a graviton exchange, the antisymmetric ones require a gauge field exchange contribution leading to an effective Lagrangian with a Chern-Simons term and a Pauli term. This has been worked out in full details in  \cite{Kovensky:2018xxa}.

~

Firstly, let us very briefly recall the derivation of the symmetric structure functions. The external states of the type IIB string theory scattering amplitude are two ten-dimensional dilatino fields (which are Neveu-Schwarz-Ramond (NS-R) fields), and two graviphotons (each being in a particular polarization state of the graviton NS-NS field). Now, we consider small values of the Bjorken variable. Using the relation between the four-dimensional Mandelstam variable $s$ and the Bjorken variable, namely: $s = -(P+q)^2 \simeq q^2/x$, small $x$ values are related to large center-of-mass energy $\sqrt{s}$. Thus, taking into account the $1/z^2$ warp factor of the metric (\ref{metric1}) the corresponding ten-dimensional Mandelstam variable $\tilde s = z^2 s$ is also large in the AdS bulk. This implies that the $\tilde{t}$-channel becomes dominant, which tells us that if a spin-$j$ particle is exchanged its contribution gives a factor ${\tilde s}^j$. In this case, the leading process implies the exchange of a Reggeized graviton, being $j \sim 2$.

The starting point now is the four-point closed-string theory scattering amplitude in ten-dimensional Minkowski space, which by virtue of the Kawai-Lewellen-Tye relations, factorizes as the product of two open-string theory scattering amplitudes as follows
\begin{equation}
{\cal{A}}(1,2,\tilde{3},\tilde{4}) = 4\ i\ \kappa_{10}^2 \,
{\cal{G}}(\alpha',\tilde{s},\tilde{t},\tilde{u}) \,
K_{op}^{\mathrm{bos}}(1,2,3,4) \otimes
K_{op}^{\mathrm{fer}}(\tilde{3},1,2,\tilde{4}) \, .
\end{equation}
In this expression the open string theory kinematic factors are denoted by $K_{op}$. Fermionic modes are indicated with tildes. 
The kinematic factor involving only bosons is given by
\begin{equation}
K_{op}^{\mathrm{bos}}(1,2,3,4) =
\xi_1^{M}\xi_2^{N}\xi_3^{P}\xi_4^{Q} \left[-1/4\,
\tilde{s}\,\tilde{u}\, \eta_{MN}\eta_{PQ} + \cdots \right] \; ,
\label{kopbos}
\end{equation}
while for two bosons and two fermions the factor is
\bea
K_{op}^{\mathrm{fer}}(\tilde{3},1,2,\tilde{4}) =
\xi_1^{M'}\xi_2^{N'}\bar{u}_3^\alpha u_4^\beta
\left[\tilde{s}\left(k^2_{M'}(\Gamma_{N'})_{\alpha\beta} -
k^1_{N'}(\Gamma_{M'})_{\alpha\beta} -
\eta_{M'N'}(\Gamma^{P})_{\alpha\beta} k^2_{P} \right) + \cdots
\right] \ . \label{kopfer} \nonumber \\
\eea
Terms leading to sub-dominant contributions in the dual DIS process are indicated with dots. $\Gamma^N$ stands for the ten-dimensional Gamma matrices. $\xi_i$ indicates polarization of bosons, while $u_i$ is used for polarization of fermions.
We use capital Latin letters for ten-dimensional bosonic indices and Greek letters $\alpha, \beta$ for spinor indices. 

The ten-dimensional Mandelstam variables are defined as
\begin{equation}
\tilde{s}=-(k_1 + k_4)^2 , \ \ \ \ \ \tilde{t}=-(k_1 + k_2)^2  \ \ \ \ \text{and} \ \ \ \ \tilde{u}=-(k_1 + k_3)^2 \, ,
\end{equation}
being $k_1$ and $k_2$ the ten-momenta of the first and second graviphotons. On the other hand, the ten-momenta of the two dilatinos are $k_3$ and $k_4$. The polarizations of the graviphotons and dilatinos are
\beq
h^{MN}_i \equiv \xi_i^M \otimes \xi_i^N \ \ \  \text{and} \ \ \quad
(\Gamma^M)_\beta^\alpha \hat{\lambda}^\beta_i \equiv u_i^\alpha \otimes \xi_i^M\, , \label{closedpolarization}
\eeq
respectively. From these expressions one obtains the following effective action, where the label (S) indicates that the symmetric structure functions can be derived from it,
\bea
S_{{\mathrm{eff}}}^{{\mathrm{(S)}}} = 2 \,\kappa_5^2\,
\textrm{Im}\left[\tilde{s}^2 \,
{\cal{G}}(\al',\tilde{s},\tilde{t},\tilde{u})\right] C\int
d^5x\sqrt{g_{AdS_5}}\ F_{mp}F_n^{\phantom{n}p}\,
\bar{\lambda}\gamma^{(m}\der^{n)}\lambda \, , \label{SheurSim}
\eea
where
\begin{equation}
\tilde{s}^2 \, {\cal{G}}(\al',\tilde{s},\tilde{t},\tilde{u}) = -
\frac{\al'^3 \tilde{s}^2}{64} \prod_{\chi =
\tilde{s},\tilde{t},\tilde{u}} \frac{\Gamma \left(-\al' \chi
/4\right)}{\Gamma\left(1+\al' \chi /4\right)} \, .
\end{equation}

Next, one evaluates the effective action (\ref{SheurSim}) on-shell, and using of the optical theorem one obtains
\begin{equation}
S_{{\mathrm{eff}}}^{{\mathrm{Sym}}} \equiv n_\mu n^*_\nu
\,{\mathrm{Im}} \, \left[T^{\mu\nu}_{(S)}\right] = \frac{1}{2\pi}
n_\mu n^*_\nu \, W^{\mu\nu}_{(S)} \, ,
\end{equation}
from which the symmetric structure functions are derived, obtaining the full symmetric structure functions for a spin-1/2 hadron
\bea
&&F_1\left(x,q^2\right) =
\frac{1}{x^2}\left(\frac{\Lambda^{2}}{q^2}\right)^{\tau-1}\frac{\pi^2|c_i'|^2
C}{4(4\pi\lambda_{\text{'t Hooft}})^{1/2}}I_{1,2\tau+3} \, , \\ \nonumber
&& \\
&&F_2\left(x,q^2\right) =  2x\frac{2\tau+3}{\tau+2}F_1(x,q^2) \, ,
\label{F222} \\
\eea
while
\bea
g_3\left(x,q^2\right)=g_4\left(x,q^2\right)=g_5\left(x,q^2\right)=0
\, ,
\eea
and
\beq
I_{j,n} = \int_0^\infty dw\ w^n K_j^2(w) =
2^{n-2}\frac{\Gamma(\n+j)\Gamma(\n-j)\Gamma(\n)^2}{\Gamma(2\n)} \ ,
\ \n=\frac{1}{2}(n+1) \ , \ I_{1,n}=\frac{n+1}{n-1}I_{0,n} \, ,
\eeq
for the Bessel functions of second kind, and the twist $\tau \equiv \Delta - s$, where $\Delta$ is the conformal dimension of the operator with spin $s$. The functions $g_3$, $g_4$ and $g_5$ are zero in this parametric region. Let us emphasize that from the $t$-channel graviton exchange there are no contributions to the antisymmetric structure functions.

~

Next, we briefly show how the antisymmetric structure
functions can be derived from type IIB superstring theory. 
In this case the holographic dual calculation is given through a gauge field exchange in the $\tilde{t}$-channel within the AdS-bulk geometry. Thus, one has to derive the effective Lagrangian from type IIB superstring theory, and then calculate the antisymmetric structure functions. The four-point closed string theory scattering amplitude must have external R-R states, since the massless gauge fields $A_m^C$ of the 5-dimensional $SU(4)$ gauged supergravity are linear combinations of two low-lying Kaluza-Klein modes on $S^5$, coming from both NS-NS (graviton $h_{MN}$) and R-R (a R-R 4-form field ${\cal{C}}_{M_1\cdots M_4}$) string states. Thus, 
\begin{equation}
{\cal{A}}(\tilde{1},\tilde{2},\textit{3},\textit{4}) = - i\ \kappa^2
{\cal{G}}(\alpha',\tilde{s},\tilde{t},\tilde{u})\,
K_{op}^{\mathrm{fer}}(\tilde{1},\tilde{2},\tilde{3},\tilde{4})\otimes
K_{op}^{\mathrm{fer}}(\tilde{3},1,2,\tilde{4}) \,,
\end{equation}
where the Italic numbers indicate R-R fields. In addition, 
\beq
K_{op}^{\mathrm{fer}}(\tilde{1},\tilde{2},\tilde{3},\tilde{4}) =
\frac{\tilde{s}}{2} \,\bar{u}_1\Gamma^{M}u_2\,\bar{u}_3\Gamma_{M}u_4
\, ,
\eeq
while the second kinematic factor is given in (\ref{kopfer}). 
The polarizations of the dilatino fields are given in equation (\ref{closedpolarization}). On the other hand, 
\beq
u_i^\alpha \otimes \bar{u}_i^\beta = ({\cal{C}}_Q
\Gamma_{i(5)})^{\alpha\beta}\,,\quad \mathrm{with}\quad
\Gamma_{i(5)}=({\cal{F}}_i)_{M_1\cdots M_5}\Gamma^{M_1\cdots M_5} \,
,
\eeq
are the polarizations of the 4-form field. ${\cal{C}}_Q$ is the
charge conjugation matrix. The leading amplitude necessary to write the effective Lagrangian becomes
\bea
{\cal{A}}(\tilde{1},\tilde{2},\tilde{3},\tilde{4}) &=& - i\ \kappa^2
{\cal{G}}(\alpha',\tilde{s},\tilde{t},\tilde{u})\, \tilde{s}^2
\frac{16}{15} ({\cal{F}}_{3})_{M M2\cdots
M_5}({\cal{F}}_4)_{N}^{\phantom{a}M_1\cdots M_5} \bar{{\hat{ \lambda}}}_1
\gamma^{(N}k_2^{M)}{\hat{\lambda}}_2 \, .
\eea

As for the symmetric structure functions, the relation between the effective on-shell action and the hadronic tensor is
\begin{equation}
-iS_{\mathrm{eff}}^{\mathrm{(A)}} \equiv n_\mu n^*_\nu
\,{\mathrm{Im}} \left[T^{\mu\nu}_{\mathrm{(A)}}\right] =
\frac{1}{2\pi} n_\mu n^*_\nu W^{\mu\nu}_{\mathrm{(A)}} \, .
\end{equation}
After the evaluation of this action,
\beq
n_\mu n^*_\nu \,{\mathrm{Im}}
\left[T^{\mu\nu}_{\mathrm{(A)}}\right] = \varepsilon^{\mu \nu \rho
\sigma} n_\m n_\n^* q_{\rho}\,P_\sigma q^{-2} {\cal{Q}}\ \frac{\pi\,
|c_i|^2}{12\sqrt{4\pi\lambda}}
\left(\frac{\Lambda^2}{q^2}\right)^{\tau-1}  {\cal{I}}_{\tau} \, ,
\label{ImSA}
\eeq
where ${\cal{Q}} \equiv d_{33C}{\cal{Q}}^C$, being $d_{33C}$ the complete symmetric symbol for the $SU(4)$ gauge group in the 5-dimensional gauged supergravity obtained from reduction of type IIB supergravity on the 5-sphere. This contribution (\ref{ImSA}) is related to the Chern-Simons term \cite{Kovensky:2018xxa}.
Then, the antisymmetric structure functions are
\beq
F_3^{CS}\left(x, q^2\right) =
\frac{1}{x}\left(\frac{\Lambda^2}{q^2}\right)^{\tau-1} \ {\cal{Q}} \ \frac{\pi^2
|c_i|^2 }{6\sqrt{4\pi\lambda_{\text{'t Hooft}}}} \ {\cal{I}}_{\tau} \, , \label{F3}
\eeq
where
\begin{equation}
{\cal{I}}_\tau \equiv \int d\omega \ \omega^{2\tau+2} \
K_0(\omega) \ K_1(\omega)= \frac{\sqrt{\pi}}{4}
\frac{\Gamma^2\left(\tau+1\right)
\Gamma\left(\tau+2\right)}{\Gamma\left(\tau+\frac{3}{2}\right)} \, ,
\end{equation}
and $g_1^{CS}\left(x,q^2\right) = g_2^{CS}\left(x,q^2\right) = 0$.

At this point it is very important to emphasize that there are examples of holographic dual models similar to ${\cal{N}}=4$ SYM in the UV, which in the IR show spontaneously broken $R$-symmetry \cite{Hatta:2009ra}. For these models our present calculation leads to
\begin{equation}
g_1^{CS}(x,q^2) = \frac{1}{2} \, F_3^{CS}(x,q^2) \propto \frac{1}{x}
\, . \label{2g1F3}
\end{equation}
We shall assume this behaviour in the present work as in the work developed in references \cite{Hatta:2009ra} and \cite{Kovensky:2018xxa}. In addition, there is a second contribution due to a Pauli (P) term in the effective Lagrangian, which is also related to the gauge field exchange in the AdS-Feynman-Witten diagram for the FCS process. For $g_1$ there is the relation between the Chern-Simons and the Pauli contributions as follows 
\bea
\frac{g_1^P}{d_{33C} \beta^C} \propto
\frac{g_1^{CS}}{d_{33C}{\cal{Q}}^C}(\tau-1) \, ,
\eea
which depends on the twist $\tau$ of the SYM theory operator. 
Therefore, at low $x$ we have $g_1=g_1^{CS}+g_1^P$.

From the string theory and supergravity point of view there are very different mechanisms responsible for the antisymmetric structure functions. In the supergravity regime $F_3$, $g_1$, $g_2$, $g_3$, $g_4$ and $g_5$, are related to the right-handed dilatino in AdS$_5$ near the boundary. At lower $x$ values, however, these functions are derived from the non-Abelian Chern-Simons and Pauli terms in the 5-dimensional effective action.

%
\subsection{The BPST Pomeron}\label{S-2-3}
%

The exponentially small $x$ region,  $ x \lesssim \exp{(-\lambda_{\text{'t Hooft}}^{1/2})}$ is described by the BPST Pomeron. Now, it is convenient to reinstate the radius $R=(4 \pi g_{\text{string}} N_c)^{1/4}$ in the AdS$_5 \times S^5$  metric. The holographic calculation that we follow holds for $N_c \gg \lambda_{\text{'t Hooft}} \gg 1$ and $g_{\text{string}} \ll 1$, being perturbative from the string theory perspective. The ambient space is described by world-sheet fields
\begin{equation}
X^M(\sigma_1, \sigma_2) = x^M + X'^M(\sigma_1, \sigma_2) \, ,
\end{equation}
where $x^M$ indicates the zero modes for each $M=0, \dots 9$. If one considers fixed zero modes, then the Gaussian integral on $X'^M$ leads to exactly the same as it would do in ten-dimensional Minkowski space. This gives the ten-dimensional flat-space $S$-matrix that would be seen by a local observer,
\begin{equation}
S = i \int d^4x \ \int d^6y \ \sqrt{-G} \ A_{\text{local}}(x, y) \, ,
\end{equation}
where it has been integrated over the zero modes.

Notice that due to the metric warp factor there is the simple but crucial red-shift
\begin{equation}
\tilde{P}^\mu_{\text{10d}} = \frac{z}{R} \ p^\mu_{\text{4d}}  \, 
\end{equation}
being $\tilde{P}^\mu_{\text{10d}}$ the inertial four-momentum measured by a local observer in the bulk, while $p^\mu_{\text{4d}}$ is the same component of the four-momentum corresponding to the gauge theory at the boundary of the AdS space. Recall that $\mu=0, 1, 2, 3$ are Minkowski four-dimensional indices.

Now, one may write $A_{\text{local}}(x, y)$ as
\begin{equation}
 A_{\text{local}}(x, y) \rightarrow \tau_{10}(\tilde{P}) \ \prod_{i=1}^m \ e^{i p_i \cdot x_i} \ \Psi(y_i) \, ,
\end{equation}
where $\tau_{10}(\tilde{P})$ is the flat-spacetime string theory scattering amplitude of $m$ external states $\Psi(y_i)$. Then, 
\begin{equation}
S = i \ (2\pi)^4 \ \delta^{(4)}(\Sigma_i p_i) \ \int d^6y \ \prod_{i=1}^m \sqrt{-G} \  \Psi(y_i) \ \tau_{10}(\tilde{P}) \, .
\end{equation}
Now, let us apply this to $2 \rightarrow 2$ particle Regge scattering. From the warp factor of the metric the red-shift leads to
\begin{equation}
\tilde{s}_{\text {10d}} = \frac{z^2}{R^2} s_{\text{4d}} \,\,\,\,\,\,\,\,\,\, \text{and} \ \ \,\,\,\,\,\,\,\,
\tilde{t}_{\text{10d}} = \frac{z^2}{R^2} t_{\text{4d}}  \, .
\end{equation}
We know that
\begin{equation}
\tau_{10}(\tilde{P}) = g_{\text{string}}^2 \ \alpha'^3 \ F(\tilde{P} \sqrt{\alpha'}) \, ,
\end{equation}
where 
\begin{equation}
F(\tilde{P} \sqrt{\alpha'}) = K(\tilde{P} \sqrt{\alpha'}) \ \left[
\prod_{\tilde{x}=\tilde{s},\tilde{t},\tilde{u}} \
\frac{\Gamma(-\alpha' \tilde{x}/4)}{\Gamma(1+\alpha' \tilde{x}/4))} \right] 
\, ,
\end{equation}
which, for $|\tilde{t}| \ll \tilde{s}$, with $\tilde s+\tilde t+\tilde u=0$, can be approximated by
\begin{equation}
F(\tilde{P} \sqrt{\alpha'})   \approx K(\tilde{P} \sqrt{\alpha'}) \ (\alpha'  \tilde{s})^{2+\alpha' \tilde{t}/2} \ \frac{\Gamma(-\alpha' \tilde{t}/4)}{\Gamma(1+\alpha' \tilde{t}/4))}  \, .
\end{equation}
Then, plugging these expressions in $\tau_{10}(\tilde{P})$
one obtains the 2 $\rightarrow$ 2 four-dimensional scattering amplitude
\begin{equation}
\tau_{4}(s, t)
 = \int d^6y \ \sqrt{-G} \ g_{\text{string}}^2 \ \alpha'^3 \ K(\tilde{P} \sqrt{\alpha'}) \ (\alpha'  \tilde{s})^{2+\alpha' \tilde{t}/2} \ \frac{\Gamma(-\alpha' \tilde{t}/4)}{\Gamma(1+\alpha' \tilde{t}/4))} \  \Pi_{\text{i=1}}^4 \  \Psi(y_i) \, .
\end{equation}
It means that the relevant exponent in the Regge limit is 
$j=2+\alpha' \tilde{t}/2=2+\alpha' t z^2/(2 R^2)$.

There are two very different physical situations. On the one hand, for positive $t$ and $0<t \ll s$, the maximum value of the exponent corresponds to the maximum value of the radial coordinate $z_0$ (recall that $0 < z < z_0$). Therefore,
\begin{equation}
j_{\text{Max}}=2+\alpha' t z_0^2/(2 R^2) \, .
\end{equation}
This is the IR region of the gauge theory. Thus, this is a non-perturbative effect for the gauge theory, related to Regge physics associated with the soft Pomeron.  On the other hand, when $t<0$ and $0<|t| \ll s$ the maximum value of the exponent is:
\begin{equation}
j_{\text{Max}}=2 \, .
\end{equation}
This is the UV region of the gauge theory, i.e. for $z\rightarrow 0$. In the gauge theory this corresponds to the (hard) BFKL Pomeron. In this way both the hard-BFKL and soft-Regge Pomerons become unified within a single holographic dual description. This is a very important result obtained in reference \cite{Brower:2006ea}. In this context, Brower, Djuric, Sarcevic and Tan \cite{Brower:2010wf} obtained the structure function $F_2$ derived from the BPST Pomeron. This function has four parameters, namely: $g_0^{2}$, $\rho$, $z_0$ and $Q'$, and it is given by
\begin{equation}
F^{\text{BPST}_{\text{HW}}}_2(x, Q^2) = \frac{g_0^2 \ \rho^{3/2} \ Q}{32 \ \pi^{5/2} \ \tau_b^{1/2} \ Q'} \ e^{(1-\rho) \tau_b} \left(e^{-\frac{\log^2{(Q/Q')}}{\rho \tau_b}}+ {\cal{F}}(x, Q, Q') \ e^{-\frac{\log^2{(Q Q' z^2_0)}}{\rho \tau_b}}\right) \, ,
    \label{FBPSTHW}
\end{equation}
where 
\begin{eqnarray}
{\cal{F}}(x, Q, Q')= 1-2 \ (\pi \ \rho \ \tau_b)^{1/2} \ e^{\eta^2(x, Q, Q')} \ {\text{erfc}}\left(\eta(x, Q, Q')\right) \, ,
\end{eqnarray}
and
\begin{eqnarray}
\eta(x, Q, Q') &=& \frac{\log{\left(z_0^2 \ Q'\ Q \right)}+ \rho \ \tau_b}{\sqrt{ \rho \ \tau_b}} \ , 
\end{eqnarray}
where
\begin{eqnarray}
\tau_b(x, Q, Q') &=& \log{ \left( \frac{\rho \ Q }{2 Q' x} \right)} \, , 
\end{eqnarray}
is a longitudinal boost. The parameter $Q' \approx 1/z'$, being $z'$ the support of the Dirac-delta distribution which approximates the hadron wave-function \cite{Brower:2010wf}. Thus, $z'$ should be of the order of the hadron size. $g_0$ is an overall constant, $\rho=2/\sqrt{\lambda_{\text{t' Hooft}}}$, and $z_0$ is the IR cut-off energy of the gauge theory. The presence of this cut-off is indicated by the label HW (hard-wall model). 

~

In addition, in reference \cite{Kovensky:2018xxa} it has been obtained the antisymmetric structure function $g_1$. This equation was obtained assuming that the kernels for $j \approx 1$ (Reggeized gauge field exchange) and $j \approx 2$ (Reggeized graviton exchange) can be approximately described in the same way \cite{Kovensky:2018xxa}. There are important changes of this derivation with respect to the derivation of the symmetric function $F_2$, since in the $\tilde{t}$-channel there is a Reggeized gauge field exchange instead of a Reggeized graviton. Thus, for instance for $t<0$ and $0<|t| \ll s$, i.e. the UV region of the gauge theory $j_{{\text Max}}=1$. The parameters $\rho,z_0, Q'$ are to be obtained from the $F_2$ fit to experimental data. Therefore, there is only one new free parameter $C$ to fit to all $g_1$ data. The corresponding expression for $g_1(x, Q^2)$ is
\begin{equation}
g^{\text{BPST}_{\text{HW}}}_1(x, Q^2)  = \frac{C \rho^{-1/2} \ e^{(1-\frac{\rho}{4}) \tau_b}}{\tau_b^{1/2}} \left(e^{-\frac{\log^2{(Q/Q')}}{\rho \tau_b}}+ {\cal{F}}(x, Q, Q') \ e^{-\frac{\log^2{(Q Q' z^2_0)}}{\rho \tau_b}}\right)  \, .
    \label{g1BPSTHW}
\end{equation}


%
\section{Comparison with experimental data for $F_2^P(x, Q^2)$}\label{S-3}
%

In this section we carry out an extensive comparison with modern experimental data from several collaborations for the proton. Most of data correspond to very small $x$ values. In this range it turns out that the dual description, in terms of the holographic Pomeron, fits data very well. Recall that in this kinematic range gluon dynamics is dominant, thus the top-down description we study is able to capture these effects. In this sense there are similarities between ${\cal {N}}=4$ SYM and QCD.

On the other hand, for larger values of the Bjorken parameter, the dual supergravity description based on ${\cal {N}}=4$ SYM theory is not good to fit experimental data. This is due to the lack of matter in the fundamental representation in this model, thus not allowing to describe valence quarks, whose physics dominates the hadron structure for this kinematic regime.

We present the results starting in section \ref{S-3-1} with the situation that fits better, i.e. the exponentially small-$x$ domain described in terms of a single BPST Pomeron exchange. Then, for the intermediate region we consider two descriptions that we explain in two subsections. In subsection \ref{S-3-2-1} we use a second single BPST Pomeron exchange, while in subsection \ref{S-3-2-1} we also add the contribution from type IIB superstring theory scattering amplitudes. For larger values of $x$, in section \ref{S-3-3} we show the results of the fit using type IIB supergravity.

%
\subsection{$F_2^P$ at low $x$ and the BPST Pomeron}\label{S-3-1}
%

In the range of the Bjorken variable $0< x <0.01$ Brower et al. have found that by using the BPST Pomeron with an IR cut-off, the structure function $F_2$ gives a remarkably good fit \cite{Brower:2010wf} in comparison with experimental data of the proton, corresponding to the H1-ZEUS collaboration \cite{H1:2009pze} of HERA small-$x$ DIS scattering experiments. In this case the values of $Q^2$ are within the range from 0.1 GeV$^2$ to 400 GeV$^2$.

In order to check the consistency with previous results, firstly we have carried out a similar fit as the one obtained by Brower et al. \cite{Brower:2010wf}. The parameters are detailed in the first line of table 1. It includes 249 experimental points, while the BPST Pomeron has four free parameters. The fit leads to $\chi^2_{\text {total}}=328$, with the value per degree of freedom $\chi^2_{\text{d.o.f.}}=1.34$ and a $P$-value 0.00031. Then, in order to improve this fit we have implemented a sieving procedure following \cite{Block:2005qm}. This allows one to make a robust fit, by excluding in a consistent way a limited number of points (considered as ``outliers") whose individual $\Delta \chi^2_i$ values are larger or equal to a certain value $\Delta \chi^2_{\text {max}}=4$. All the information about $\Delta \chi^2_{\text{max}}$ together with the total number of points included in each fit for all cases are displayed in table 1 (and table 2 for $g_1$). Technical details of the sieving procedure are described in the appendix. Also, some figures showing $\chi^2_{\text{d.o.f.}}$ for different values of $\Delta \chi^2_{\text {max}}$ are presented.

We take the following definitions: the $\chi^2$ per degree of freedom is given by
%
%
$\chi^2_{\text{d.o.f.}} = \frac{\chi^2_{\text{total}}}{N_{\text{d.o.f.}}} $,
%
%
where $N_{\text{d.o.f.}}$ is the difference between the number $N_p$ of experimental points included in the fit and the number of parameters, with
\beq
\chi^2_{\text{total}} = \sum_{i=1}^{N_p} \Delta\chi^2_i \, ,
\eeq
being the $\Delta\chi^2_i$ defined in equation (\ref{DeltaChi2}) in the appendix. The $P$-value is defined as follows,
\beq
P(\chi^2_{\text{total}}, N_{\text{d.o.f.}}) = \frac{1}{2^{N_{\text{d.o.f.}}/2} \ \Gamma[N_{\text{d.o.f.}}/2]} \ \int_{\chi^2_{\text{total}}}^\infty \ t^{\frac{N_{\text{d.o.f.}}}{2}-1}  \  e^{-t/2} \, dt \, .
\eeq

\begin{figure}[H]
\centering
\includegraphics[scale=0.8]{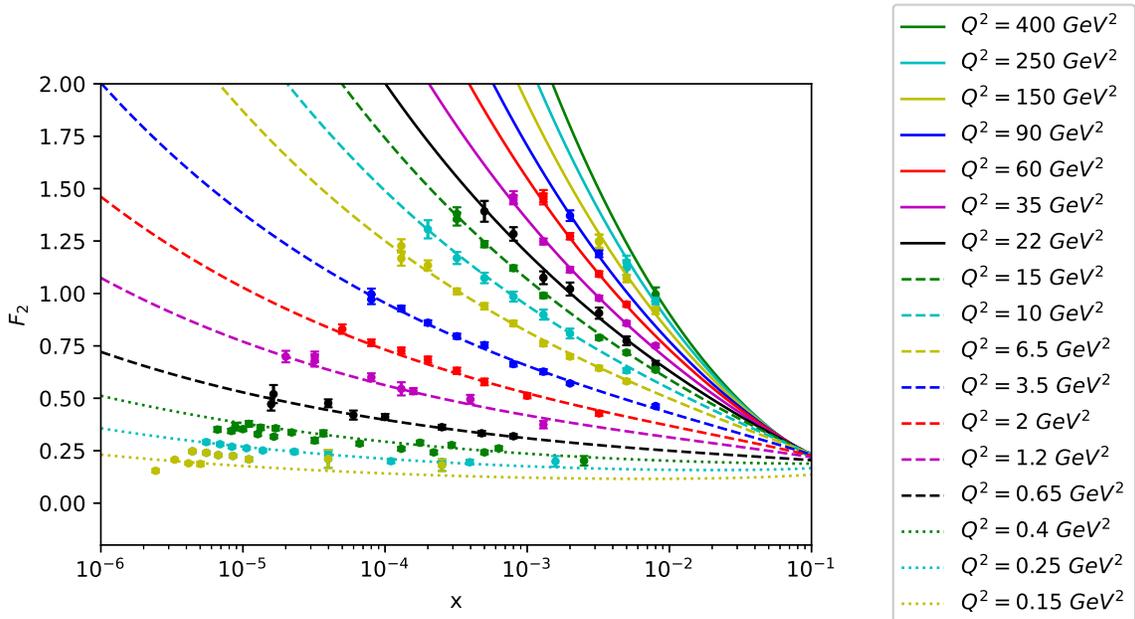}
\caption{{\small Best fit for $F_2$ obtained from the hard-wall BPST Pomeron in comparison exclusively with H1-ZEUS data for the proton corresponding to low-$x$ DIS at HERA. Other collaborations are included in others figures. From the initial 249 points in the ranges  $x<10^{-2}$ and  0.1 GeV$^2 < Q^2 \leq 400$ GeV$^2$, using a sieving with $\Delta\chi^2_{\text {max}}=4$ there are 228 experimental points left, leading to $\chi^2_{\text{d.o.f.}}=1.07$. This reproduces the results of Brower et al. \cite{Brower:2010wf}. We only display 17 curves for certain representative values in the above range of $Q^2$ as indicated in the box at the right (in fact these are the same values of reference \cite{Brower:2010wf}, chosen to compare with it). Also notice that although the horizontal axis includes a range beyond $x=0.01$, i.e. up to $x=0.1$, in this figure we do not display experimental points for $x>0.01$. The horizontal axis has a logarithmic scale. Error bars of data are shown for each point. In some cases dots representing data points are larger than the corresponding error bars.}}
\label{Fig_h1n1_x_sx}
\end{figure}

$F_2$ as a function of the Bjorken parameter is depicted in figure \ref{Fig_h1n1_x_sx} for 17 values of $Q^2$. In this case points with $\Delta\chi^2_i \geq \Delta\chi^2_{\text {max}}=4$ have been excluded, leading to a total of 228 points and $\chi^2_{\text{d.o.f.}}=1.07$, with $\chi^2_{\text{total}}=240$ and $P=0.22066$. This very good fit, similar to the one obtained in reference \cite{Brower:2010wf}, motivates us to investigate this structure function including more data and extending the kinematic range. This fit is given in the second line of table 1.

\begin{center}
\begin{tabular}{ |c|l|c|c|c|c|c|c|c|c| }
\hline
& {\small Model} & $x$ & $N_p$& sieving & $\rho =$ & $g^2_0$ & $z_0$ & $Q'$ & $\chi^2_{\text{d.o.f.}}$ \\
&       & range & exp. & $\Delta \chi^2_{max}$ &     $2\lambda_{\text{'t Hooft}}^{-1/2}$     &       & [GeV$^{-1}$]     &  [GeV]    & \\
\hline \hline

1 & {\small hard-wall}  & $<0.01$ & 249    & No &0.7776  &105.01  &5.039  &0.4632  & 1.34 \\
& {\small BPST}    &         &     &    &$\pm 0.0019$  & $\pm0.85$ &$\pm 0.076$  & $\pm 0.0122$   &\\
\hline
2 & {\small hard-wall}  & $<0.01$ & 228    & 4  &0.7791  &103.14  &4.959  &0.4332 & 1.07 \\
& {\small BPST}    &         &     &    &$\pm 0.0016$  & $\pm0.798$ &$\pm 0.062$  & $\pm 0.0115$    &\\
\hline

3 & {\small hard-wall}  & $<0.01$ & 305    & No &0.7743  &105.42  &5.0104  &0.4838  & 1.28 \\
& {\small BPST}              &         &     &    &$\pm 0.0016$  & $\pm0.80$ &$\pm 0.0741$  & $\pm 0.0099$    &\\
\hline
4 & {\small hard-wall}  & $<0.01$ & 280    & 4 &0.7729  &103.73  &4.894 &0.4715  & 1.086 \\
& {\small BPST}              &         &     &    &$\pm 0.0014$  & $\pm0.757$ &$\pm 0.061$  & $\pm 0.0093$   &\\
\hline

5 & {\small hard-wall}  & $<0.1$ &548  & No &0.8314  &139.25  &10.57  & 0.5400  &12.08 \\
& {\small BPST}         &         &     &    &$\pm 0.003 $  & $\pm1.12 $ & $\pm0.99 $ & $\pm0.015 $   &\\
\hline
6 & {\small hard-wall} & [0.01, 0.1] & 243 & No &0.9176  &158.06  & 3.903 &0.5012  & 2.23 \\
& {\small BPST}          &         &     &    &$\pm0.0037  $  &$\pm1.06  $  &$\pm0.298 $  &$\pm0.0265 $    &\\
\hline
7 & {\small hard-wall} & [0.01, 0.1] & 201 & 4 &0.9194  &157.96  &3.751  &0.4782   & 1.25 \\
& {\small BPST}     &         &     &  & $\pm 0.0032$&$\pm 0.80$  &$\pm 0.205$  & $\pm0.0307$  & \\
\hline
8 & {\small BPST} & [0.01, 0.1] & 243 & No &0.9135  &157.90  &3.537  &0.4765  & 2.18 \\
& {\small $\oplus$ str. th.}           &         &     &  &$\pm0.0039 $  & $\pm 0.83 $ &$\pm0.250 $  &$\pm0.0536 $  & \\
\hline
9 & {\small BPST}   & [0.01, 0.1] & 204 & 4 &0.9165  &159.64  &3.477  &0.4049 & 1.24 \\
& {\small $\oplus$ str. th.}         &         &     &  &$\pm 0.0028$  & $\pm 7.11$  & $\pm 0.065$  &$\pm 0.1691$  &  \\
\hline
\end{tabular}
\vspace{0.5cm}
\end{center}
{\small Table 1. Main results of the present work for different  fits of the proton structure function $F_2(x, Q^2)$. In lines 8 and 9 the values of the constant $C_{\text{st}}$, corresponding to the contribution from the string theory scattering amplitude to the linear combination with the hard-wall BPST Pomeron, are $9 \times 10^{-5} \pm 4 \times 10^{-5}$ and $9 \times 10^{-5} \pm 3 \times 10^{-5}$, respectively. $N_p$ represents the number of experimental points in each fit. In lines 1 to 7 there are 4 parameters, in lines 8 and 9 there are 5 parameters. \footnote{Notice that the central values for all parameters shown in lines 1 and 2 coincide with their corresponding central values of the fits obtained in \cite{Brower:2010wf}. However, their corresponding errors are slightly different in comparison with the ones shown in that reference. We have checked it carefully and conclude that our results presented table 1 are correct.}}

~

\begin{figure}[H]
\centering
\includegraphics[scale=0.8]{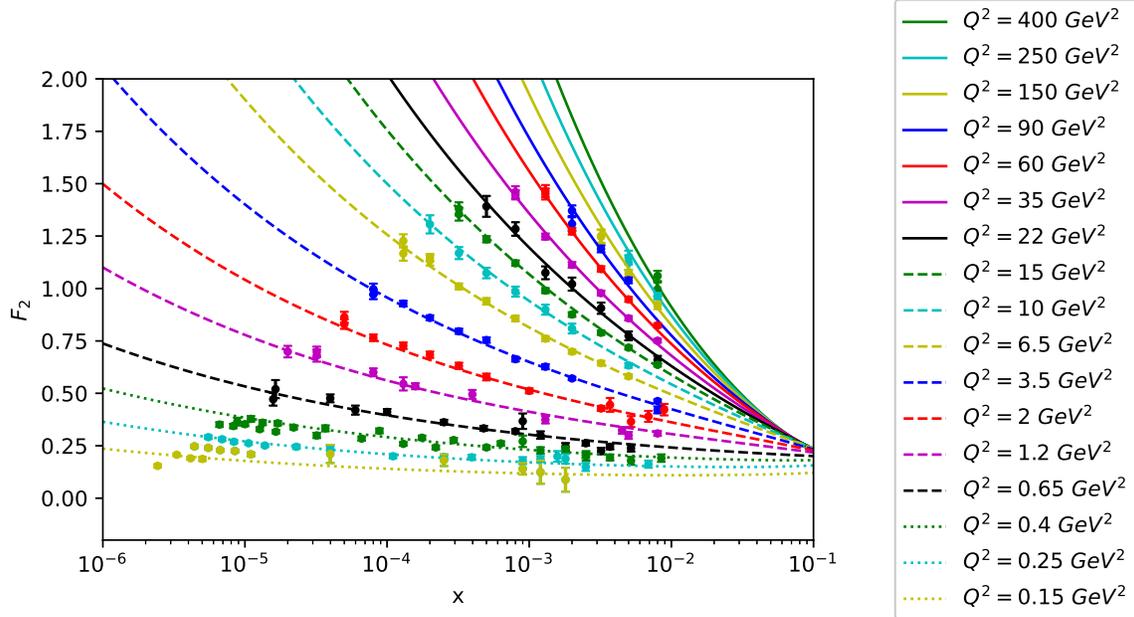}
\caption[ ]
{\small $F_2$ structure function using a single BPST Pomeron exchange to fit data of H1-ZEUS collaboration \cite{H1:2015ubc}, as well as data from BCDMS \cite{BCDMS:1989qop}, NMC \cite{NewMuon:1996fwh}, E665 \cite{E665:1996mob} and SLAC \cite{Whitlow:1991uw} collaborations within the ranges 0.1 GeV$^2< Q^2 \leq 400$ GeV$^2$ and $0<x<0.01$, corresponding to the proton. The horizontal scale is $\log x$. The number of  experimental points depicted has been limited in order to be able to visualize how a few curves fit the data. Error bars are indicated. In total the fit includes 280 data points, while $\chi^2_{\text{d.o.f.}}$ is now 1.086. The same applies to figure \ref{Fig_compl_q_sx.eps}.} 
        \label{Fig_compl_x_sx.eps}
    \end{figure}

\begin{figure}[H]
\centering
\includegraphics[scale=0.8]{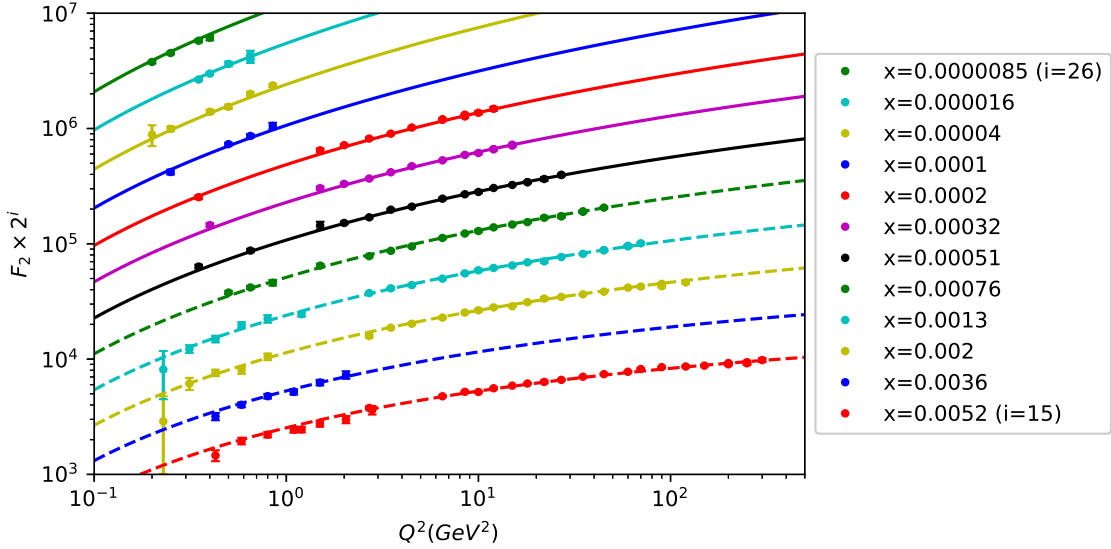}
\caption[ ]
{\small Curves for the fit of $F_2$ structure function as a function of $\log Q^2$ using to the BPST Pomeron are drawn in comparison with data from H1-ZEUS \cite{H1:2015ubc}, BCDMS \cite{BCDMS:1989qop}, NMC \cite{NewMuon:1996fwh}, E665 \cite{E665:1996mob} and SLAC \cite{Whitlow:1991uw} collaborations. Notice that the values of $F_2^P$ have been multiplied by $2^{i_x}$, where $i_x$ is the number of the $x$-bin, ranging from $i_x = 15$ $(x = 0.0052)$ to $i_x = 26$ $(x = 0.0000085)$. The range of $Q^2$ goes from 0.1 GeV$^2$ to 400 GeV$^2$.} 
        \label{Fig_compl_q_sx.eps}
    \end{figure}

Next, in the same range of the Bjorken parameter we include data of other experimental collaborations which increases the total number of points to 305 (before sieving). We consider a more recent paper of H1-ZEUS collaboration \cite{H1:2015ubc}, as well as data from BCDMS collaboration \cite{BCDMS:1989qop}, NMC collaboration \cite{NewMuon:1996fwh}, E665 collaboration \cite{E665:1996mob} and from SLAC collaboration \cite{Whitlow:1991uw}. What is interesting now is the possibility of dealing with more experimental points, which can be seen by visual inspection of figure \ref{Fig_compl_x_sx.eps}, in the range of $0.001 < x < 0.01$ and for $Q^2 < 6.5$ GeV$^2$, in comparison with figure \ref{Fig_h1n1_x_sx}. In this case the hard-wall BPST Pomeron fits the whole set of 305 points leading to a normalized $\chi^2_{\text{d.o.f.}}=1.28$ (in this case $\chi^2_{\text{total}}=383$ and $P=0.00095$). The corresponding set of parameters is presented in the third line of table 1. Then, carrying out a sieving with $\Delta\chi^2_{\text{max}}=4$, there are 280 points left and the corresponding normalized $\chi^2_{\text{d.o.f.}}$ gives 1.086, which is still very good with the addition of having 30 more points than in reference \cite{Brower:2010wf} (now we obtain $\chi^2_{\text{total}}=300$ and $P=0.1535$). This is shown in line 4 of table 1. In figure \ref{Fig_compl_x_sx.eps} we show $F_2$ as a function of the Bjorken variable for different values of $Q^2$. In addition, figure \ref{Fig_compl_q_sx.eps} displays $F_2\times 2^i$ as a function of $Q^2$ for different values of $x$. The integer $i$ is indicated in this figure. This factor is included to facilitate the visualization of the curves. 

\begin{figure}[H]
\centering
\includegraphics[scale=0.8]{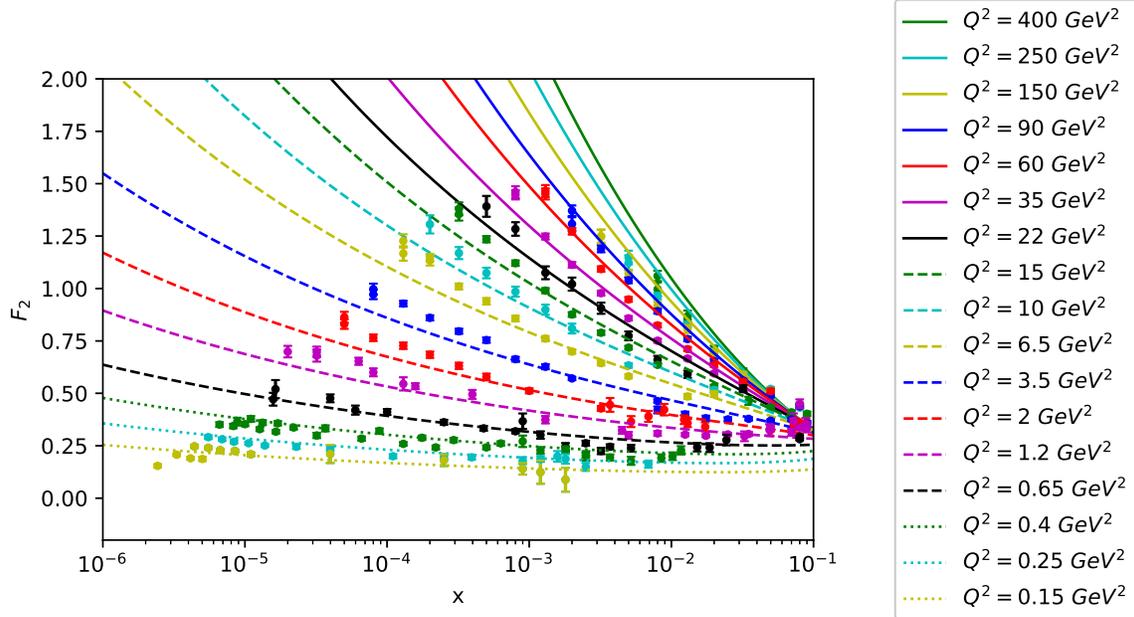}
\caption{{\small Best fit of the structure function $F_2$ from a single hard-wall BPST Pomeron to the H1-ZEUS \cite{H1:2015ubc}, BCDMS \cite{BCDMS:1989qop}, NMC \cite{NewMuon:1996fwh}, E665 \cite{E665:1996mob} and SLAC \cite{Whitlow:1991uw} experimental points at 0.1 GeV$^2 < Q^2 \leq 400$ GeV$^2$ and the extended range $0<x<0.1$, corresponding to the proton. In total there are 548 points. $\chi^2_{\text{d.o.f.}}$ = 12.08 which indicates that the fit does not reflect accurately the experimental results, particularly for $0.01 < x < 0.1$.}} 
\label{Fig_compl_x_smx}
\end{figure}

A natural question is now what happens if we try to extend the range of the Bjorken variable beyond $x \sim 0.01$. Since there is a reasonable amount of data (548 points in total) we may try for instance to consider a wider range such as $0 < x < 0.1$. Using the hard-wall BPST Pomeron of equation (\ref{FBPSTHW}) we find that the normalized $\chi^2_{\text{d.o.f.}} = 12.08$, indicating that this particular fit does not work. The parameters are presented in the fifth line of table 1. In addition, after using the sieving method the fit does not improve. This can be seen from figure \ref{Fig_compl_x_smx}, by looking closely at the region $0.01 < x < 0.1$ where we can very easily see how the curves do not fit well the experimental data. Also, in this case one can observe how the curves slightly depart from the experimental points almost everywhere.
It is interesting to notice that although the functional form of $F_2$ derived from the hard-wall BPST Pomeron shows a similar trend as shown by the experimental data around $x \sim 0.14$, where it appears a pivotal point, it is not able to fit the whole range $0<x<0.1$ with a reasonable value of $\chi^2_{\text{d.o.f.}}$.
From this we conclude that there are some effects that a single holographic Pomeron cannot capture if the $x$ range is extended from very low values towards moderately low values of $x$. This behaviour makes sense if we keep in mind that the construction of the BPST Pomeron in principle attains to the exponentially small-$x$ values, while for moderately small-$x$ values the holographic dual top-down construction has been done in terms of type IIB string theory scattering amplitudes. On the other hand, the approach based on string theory scattering amplitudes leads to a behaviour of $F_2$ proportional to the inverse power of the square of the virtual photon momentum transfer. This effect of decreasing $F_2$ as $Q^2$ increases for $x$ fixed does not match the experimental data, which in turn show an increasing trend for $F_2$ with $Q^2$ for $x$ fixed. This is a motivation for considering the possibility of a combination of the contributions from the string theory scattering amplitudes and a BPST-Pomeron exchange for the intermediate range $0.01<x<0.1$. We implement it as an effective top-down holographic dual description of the data. All this suggests several possible directions to investigate that we describe below.

%
\subsection{$F_2^P$ at intermediate $x$ and type IIB superstring theory}\label{S-3-2}
%

From the results of the previous subsection we may conclude that there is no way to fit reasonably well the set of experimental data for an extended range of the Bjorken parameter like $0 < x < 0.1$ using a unique single BPST Pomeron in this whole range. Therefore, since the parametric region $0 < x < 0.01$ is very well represented by a single hard-wall BPST-Pomeron exchange, now we focus on the range $0.01 < x < 0.1$. For this region one would have expected that the expression for $F_2$ derived from string theory scattering amplitudes works well. However, due to the $Q^2$ dependence it is clear that it cannot describe the data, because it shows the opposite trend at fixed values of the Bjorken parameter. Thus, within the top-down holographic dual approach that we are studying we propose to investigate two different possibilities. They are effective descriptions which we describe below. In the range $0.01 < x < 0.1$ we initially consider 243 points from  H1-ZEUS \cite{H1:2015ubc}, BCDMS \cite{BCDMS:1989qop}, NMC \cite{NewMuon:1996fwh}, E665 \cite{E665:1996mob} and SLAC \cite{Whitlow:1991uw} collaborations.

%
\subsubsection{A single hard-wall BPST-Pomeron exchange for $0.01<x<0.1$}\label{S-3-2-1}
%

\begin{figure}[H]
\centering
\includegraphics[scale=0.8]{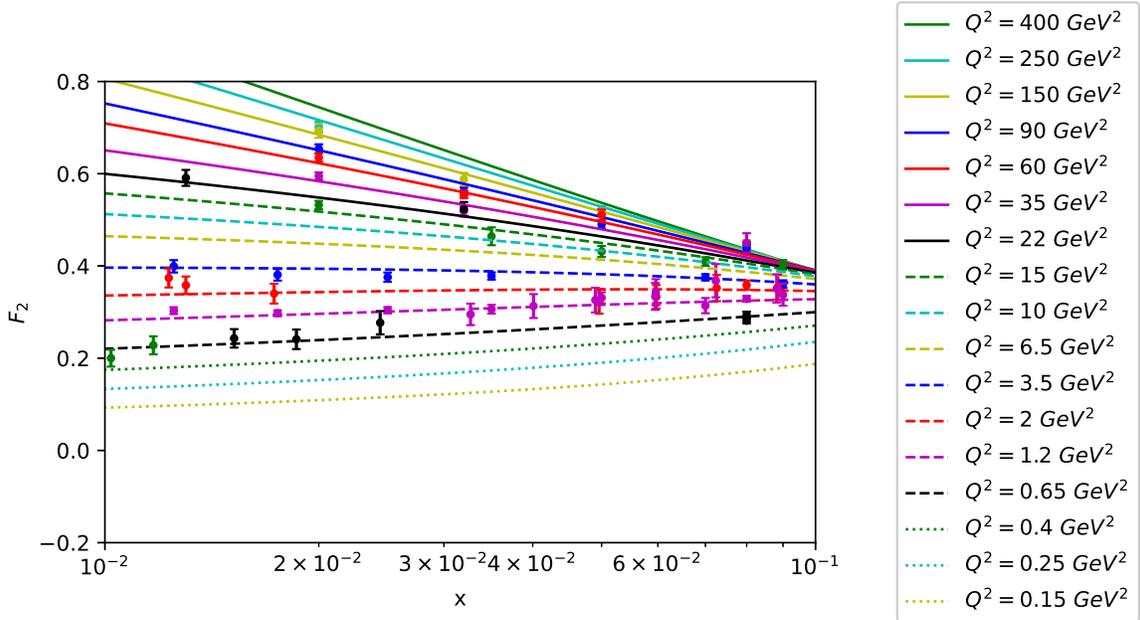}
\caption[ ]
{\small Best fit of the $F_2$ structure function from the BPST Pomeron to the H1+HERA \cite{H1:2015ubc}, BCDMS \cite{BCDMS:1989qop}, NMC \cite{NewMuon:1996fwh}, E665 \cite{E665:1996mob} and SLAC \cite{Whitlow:1991uw} collaborations, for data within the ranges 0.1 GeV$^2 < Q^2 \leq 400$ GeV$^2$ and $0.01<x<0.1$, corresponding to the proton. The horizontal scale is $\log x$. As in the previous figures the number of  experimental points depicted has being limited in order to be able to visualize how a few curves fit the data. Error bars are indicated. The same applies to figure \ref{Fig_compl_q_mx.eps}.} 
        \label{Fig_compl_x_mx.eps}
    \end{figure}

\begin{figure}[H]
\centering
\includegraphics[scale=0.8]{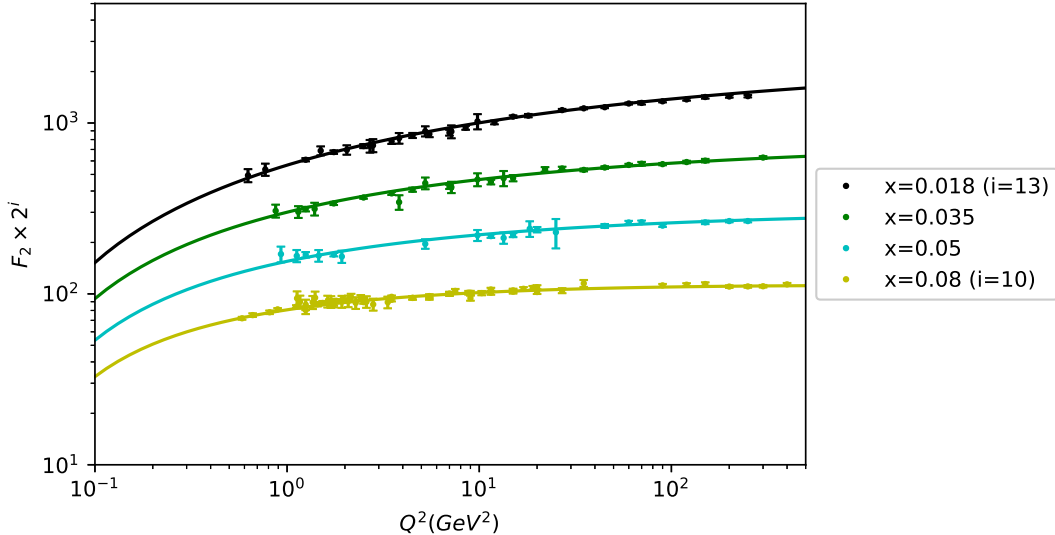}
\caption[ ]
{\small  $F_2$ as a function of $\log Q^2$, see caption of figure \ref{Fig_compl_x_mx.eps}. $F_2^P$ has been multiplied by $2^{i_x}$, where $i_x$ is the number of the $x$-bin, which goes from $i_x = 10$ $(x = 0.08)$ to $i_x = 13$ $(x = 0.018)$.}
        \label{Fig_compl_q_mx.eps}
    \end{figure}

Let us consider the expression for $F_2$ obtained from the hard-wall BPST Pomeron, but now with this equation let us fit the data restricted to the range $0.01 < x < 0.1$. Thus, effectively we have two different fits with the BPST Pomeron equation. The first one corresponds to line 4 of table 1, in the range $0 < x < 0.01$. The second fit with another BPST Pomeron is carried out for $0.01 < x < 0.1$ and the results for the 4 parameters is given in line 6 of table 1, where $\chi^2_{\text{d.o.f.}}$ is 2.23 ($\chi^2_{\text{total}}=533$ and $N_{d.o.f.}=239$) which is not good. Then, using the sieving method with $\Delta\chi^2_{\text{max}}=4$, the number of data reduces to 201, but now the normalized $\chi^2_{\text{d.o.f.}}=1.25$ indicates a better fit (see line 7 of table 1, and also we obtain $\chi^2_{\text{total}}=247$, $N_{d.o.f.}=197$ and $P=0.009$). Figure \ref{Fig_compl_x_mx.eps} displays the corresponding fit of $F_2$ as a function of $x$, and figure \ref{Fig_compl_q_mx.eps} as a function of $Q^2$, both in logarithmic scales for the horizontal axis. 
We can observe how the behaviour of this fit using a hard-wall BPST Pomeron approaches the pivotal point near $x \sim 0.1$. Also, notice that curves corresponding to the values $Q^2=$0.15, 0.25 and 0.4 GeV$^2$ in figure \ref{Fig_compl_x_mx.eps} do not display experimental points since there are not data for these values of the square virtual-photon momentum transfer, being these curves predictions of the present description. The idea behind the use of two BPST Pomerons, i.e. one for the exponentially small-$x$ region and a second one for the moderately small-$x$ values, as we just have done, relies on the fact that in the derivation of the BPST Pomeron, there is a direct relation between the parameter $\rho$ and the 't Hooft coupling: $\rho \equiv 2/\sqrt{\lambda_{\text{'t Hooft}}}$. Recall that in QCD the coupling evolves. Therefore, one may expect that the parameter $\rho$ also evolves with $Q^2$. This suggests a $Q^2$-dependence of the structure functions derived from the BPST Pomeron. On the other hand, the expressions we use do not carry such a dependence, then a single BPST Pomeron exchange with a fix value of $\rho$ is expected not to be able to reproduce data in a range beyond exponentially small values of $x$. However, the second BPST Pomeron somehow does the job, though in a very limited way, that a continuous functional $Q^2$-dependence would do.

%
\subsubsection{Combined hard-wall BPST Pomeron and string scattering amplitude}\label{S-3-2-2}
%

Next, we consider a linear combination of a contribution to $F_2$ obtained from a hard-wall BPST Pomeron exchange and a contribution from string theory scattering amplitudes. The derivation of $F_2$ associated with the string theory scattering amplitude for spin-1/2 fermions was obtained in reference \cite{Kovensky:2018xxa}, and for twist $\tau=3$ operators\footnote{Three is the lowest twist corresponding to fermionic operators of ${\cal {N}}=4$ SYM theory of the form  ${\cal {O}}_0^{(6)}(y) = C^{(6)} \, \text{Tr}(F_+ \lambda_{{\cal {N}}=4})(y)$. Its twist is $\tau=\Delta-s=3$ as explained in Section \ref{S-2}. This behaviour corresponds to the strongly coupled regime of the SYM theory. The weakly coupled gauge theory is treated perturbatively, and the OPE is dominated by twist-2 operators.} in the gauge theory this is given by
\begin{equation}
F_2^{{\text{Strings}}_{\tau=3}}(x,Q^2) = C_{\text{st}}  \  
\frac{1}{x} \left( \frac{\Lambda^2}{Q^2} \right)^2 \ ,
\end{equation}
where the constant $C_{\text{st}}$ depends on $\tau$ and on the normalization constants of the bulk fields wave-functions. 
The functional form of $F_2$ that we use for the fit is of the form
\begin{eqnarray}
F_2^{{\text{BPST}}_{\text{HW}}+{\text{Strings}}_{\tau=3}}(x, Q^2)=F_2^{{\text{BPST}}_{\text{HW}}}(x, Q^2) + C_{\text{st}} \  
\frac{1}{x} \left( \frac{1}{Q^2} \right)^2 \, . \nonumber \\ \label{HWBPSTandST}
\end{eqnarray}
This effective model can be understood as follows. A single BPST Pomeron exchange in this regime corresponds to a Reggeized graviton exchange in the type IIB string theory framework. On the other hand, the string theory scattering amplitude, which for the DIS limit in which we are interested can be obtained in terms of an effective supergravity Lagrangian through the calculation of a $\tilde{t}$-channel Feynman-Witten diagram, corresponds to the exchange of a single graviton. Therefore, in terms of the optical theorem the DIS cross section is obtained from the sum of the contribution of a Reggeized graviton (a BPST Pomeron) plus a single graviton exchange $\tilde{t}$-channel contribution. The relative normalization constant is fixed by fitting the expression (\ref{HWBPSTandST}) to the experimental data, and we will shortly see that it is very small. As an inspiration for these two contributions to DIS one may recall for instance a FCS related to inclusive processes, described in terms of the Regge theory (for a review see \cite{Collins:1977jy}). Also, although the meaning is quite different, concerning the proton+proton and proton+anti-proton hard-scattering total cross sections above the resonance region, Donanchie and Landshoff \cite{Donnachie:1992ny} and Cudell et al. \cite{Cudell:1997zi} proposed a model with two types of contributions, namely: a single Pomeron (P) exchange and a Regge (R) exchange, of the form
\begin{equation}
\sigma^{\text{total}}(s) = A^P \ s^{\alpha_P(0)-1} + A^R \ s^{\alpha_R(0)-1} \, ,
\end{equation}
where the constants $A^P$ and $A^R$ depend on each process. In particular, Donnachie and Landshoff \cite{Donnachie:1993it} predicted the following expression for $F_2$ 
\begin{equation}
F_2 (x, Q^2) = A \ x^{1-\alpha_P(0)} \ \left(\frac{Q^2}{Q^2+a}\right)^{\alpha_P(0)} + B \ x^{1-\alpha_R(0)} \ \left(\frac{Q^2}{Q^2+b}\right)^{\alpha_R(0)} \, , \label{Donnachie}
\end{equation}
with certain constraints for the constants involved, which ensure the fit to real photo-production data when $Q^2$ vanishes. In this model, the important prediction is the behaviour of $F_2$ as the Bjorken variable goes to zero, which leads to a functional form proportional to $x^{1-\alpha_P(0)} \simeq x^{0.08}$ \cite{Devenish:2004pb}. Thus, for $F_2$ the leading contribution comes from the Pomeron exchange for extremely small $x$, however, there is second contribution for larger values of $x$. The expression proposed by Donnachie and Landshoff \cite{Donnachie:1993it}, however, cannot be extrapolated for extremely small values of the Bjorken parameter, since equation (\ref{Donnachie}) does not fit the experimental data when extrapolating from intermediate values of $x$ towards very small values.

By considering equation (\ref{HWBPSTandST}), we obtain $\chi^2_{\text{d.o.f}}=2.18$, with the values of the BPST Pomeron parameters indicated in line 8 of table 1. In addition, $C_{\text{st}}=9 \times 10^{-5} \pm 4 \times 10^{-5}$. Thus, we observe that there is a very small contribution from the string theory scattering amplitude and the absolute error of $C_{\text{st}}$ is large (also notice that $\chi^2_{\text{total}}=519$ and $N_p=238$). Then, we can implement the sieving method which leads to 204 points left, while $\chi^2_{\text{d.o.f.}}=1.24$ (see line 9 of table 1), which is very close to the case discussed in the previous subsection with a hard-wall BPST Pomeron in this specific range. Now, $\chi^2_{\text{total}}=247$, $N_p=199$ and $P=0.01164$. In conclusion there is not a significant improvement of the fit in comparison with the hard-wall BPST Pomeron contribution shown in line 7.
\begin{figure}[H]
\centering
\includegraphics[scale=0.8]{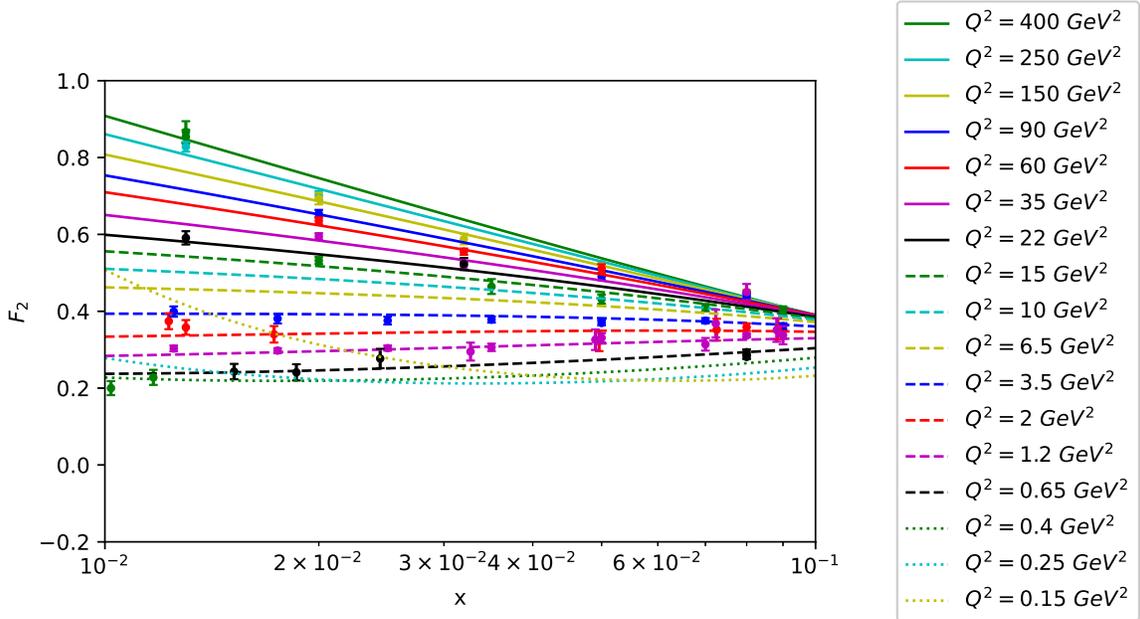}
\caption[ ]
{\small This figure displays the fit of the $F_2$ structure function from the BPST Pomeron to the H1+HERA \cite{H1:2015ubc}, BCDMS \cite{BCDMS:1989qop}, NMC \cite{NewMuon:1996fwh}, E665 \cite{E665:1996mob} and SLAC \cite{Whitlow:1991uw} data within the ranges 0.1 GeV$^2< Q^2 \leq 400$ GeV$^2$, corresponding to the proton. The horizontal scale is $\log x$. The number of  experimental points depicted has being limited in order to be able to visualize how a few curves fit the data. Error bars are indicated in both figures.} 
        \label{Fig_compl_x_mx_string.eps}
    \end{figure}
\begin{figure}[H]
\centering
\includegraphics[scale=0.8]{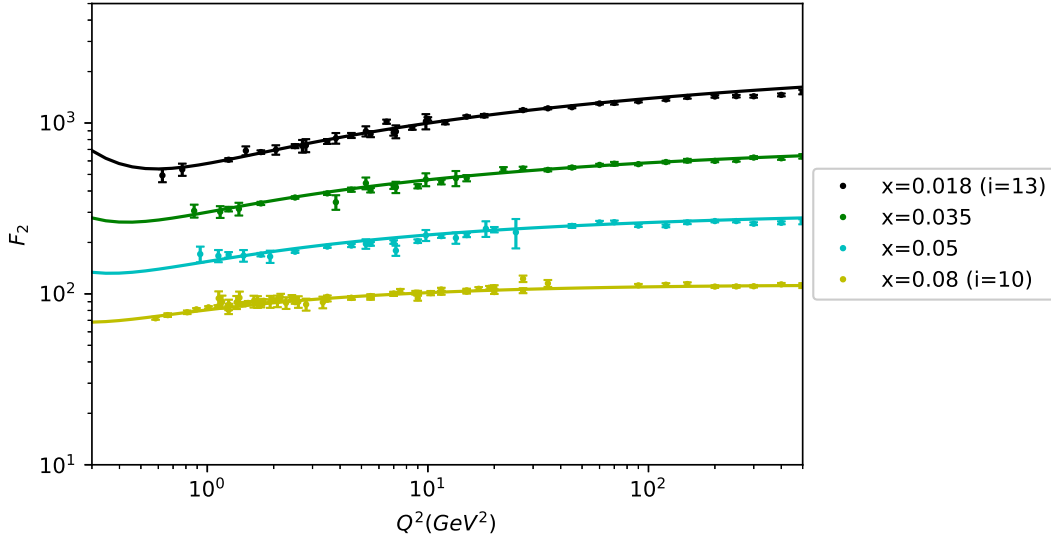}
\caption[ ]
{\small Best fit of the $F_2$ structure function from the BPST Pomeron to the H1+HERA \cite{H1:2015ubc}, BCDMS \cite{BCDMS:1989qop}, NMC \cite{NewMuon:1996fwh}, E665 \cite{E665:1996mob} and SLAC \cite{Whitlow:1991uw} data for $0.01 < x < 0.1$. The horizontal scale is  $\log Q^2$. $F_2^P$ has been multiplied by $2^{i_x}$, where $i_x$ is the number of the $x$-bin, ranging from $i_x = 10$ $(x = 0.08)$ to $i_x = 13$ $(x = 0.018)$.} 
        \label{Fig_F2_q_mx_string.eps}
    \end{figure}
For $Q^2=$ 0.15 GeV$^2$ and $Q^2=$ 0.25 GeV$^2$ we have not found experimental points. In this region, due to the inverse power of $Q^2$, the string theory contribution dominates compared with the BPST Pomeron one. Thus, these two curves have a rise towards $x \rightarrow 0.01$. These feature is not shown for larger values of $Q^2 \geq 0.4$ GeV$^2$, where the BPST Pomeron dominates. This effect becomes clearer by comparison of figures 6 and 8, where while in figure 6 there is a monotonic behaviour along all the curves, in figure 8 we can observe a different behaviour.

%
\subsection{$F_2^P$ at $0.1<x<1$ from type IIB supergravity}\label{S-3-3}
%

In section \ref{S-2-1} we have briefly described how to derive the hadron structure functions in terms of type IIB supergravity on AdS$_5 \times S^5$. In this section we present the results of the corresponding fit for the region $0.1<x<1$ for $F_2^P$.

\begin{figure}[H]
\centering
\includegraphics[scale=0.8]{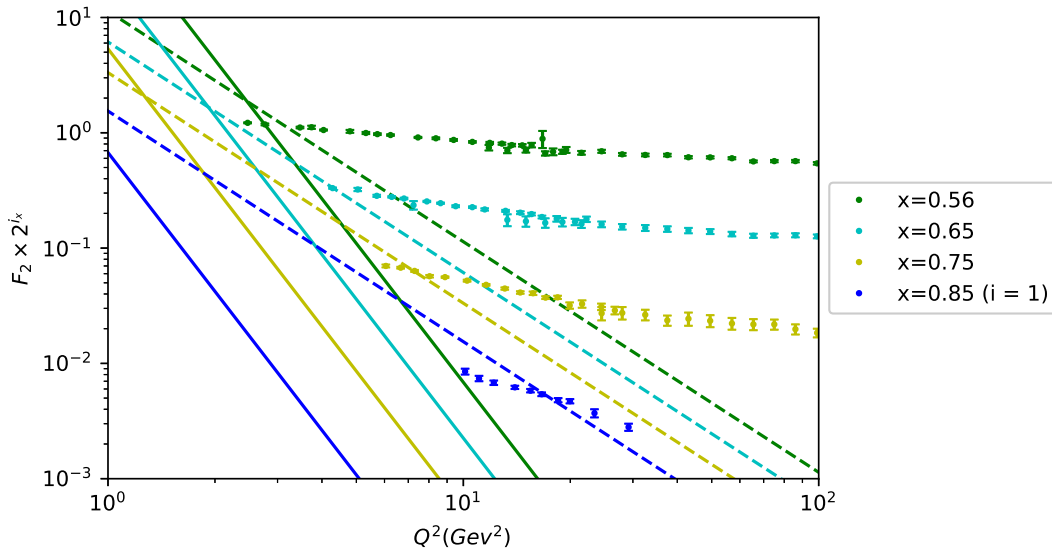}
\caption[ ]
{\small This figure displays the fit of the $F_2$ structure function from the type IIB supergravity description, within the ranges 1 GeV$^2< Q^2 \leq 100$ GeV$^2$, corresponding to the proton. Data are taken from the BCDMS \cite{BCDMS:1989qop} and SLAC \cite{Whitlow:1991uw} collaborations.} 
        \label{Fig_F2_large_x.eps}
    \end{figure}

In figure \ref{Fig_F2_large_x.eps} we show the fit of the $F_2$ structure function, considering the type IIB supergravity dual description, to data from the  BCDMS \cite{BCDMS:1989qop} and SLAC \cite{Whitlow:1991uw} collaborations. It includes data within the range 1 GeV$^2< Q^2 \leq 100$ GeV$^2$, corresponding to the proton. This is presented to illustrate how bad the dual type IIB supergravity description works in this case. Experimental points depicted as dots with their corresponding error bars cannot be fitted with the supergravity dual model. The reason why the holographic dual description fails to describe data is due to the fact that for $0.1 < x < 1$ the quark anti-quark sea becomes very important, while the particular supergravity dual model we consider does not contain quarks in the fundamental representation. This is consistent with the expectations of this model. We will discuss more on these issues in section \ref{S-6}.
The dashed lines presented in figure \ref{Fig_F2_large_x.eps} correspond to the supergravity dual description with twist-3 operators of ${\cal {N}}=4$ SYM theory. Continuous lines represent the fit using twist-5 operators corresponding to the Kaluza-Klein state $(2, 2, 2, 2, 2)$. We have shown the results for these twists and particular Kaluza-Klein states to show an example of a more general effect, which cannot be improved within the dual supergravity description.

%
\section{Antisymmetric structure function $g_1^P(x, Q^2)$}\label{S-4}
%

In this section we focus on the comparison of the antisymmetric function $g_1(x, Q^2)$ with the results of several experimental collaborations measuring properties of polarized electromagnetic DIS of the proton. As in the case of $F_2(x, Q^2)$ developed in section \ref{S-3}, we firstly consider the range $0 < x < 0.01$ for the exchange of a single holographic Pomeron. Then, we study the range $0.01 < x < 0.1$. For the antisymmetric structure function the single holographic Pomeron exchange corresponds to a Reggeized gauge field in the bulk. This is a significant difference with respect to the BPST Pomeron used for studying $F_2$.

%
\subsection{$g_1^P$ at low $x$ from the holographic Pomeron}\label{S-4-1}
%

In section \ref{S-3-1} we have investigated the fit of the BPST Pomeron for $F_2^P$ in the range $0 < x < 0.01$, which gives the values of the parameters $\rho$, $z_0$, $Q'$ and the overall constant $g^2_0$. Motivated by the very good results obtained for $F_2^P(x, Q^2)$ within this range, we now fit the holographic Pomeron associated with the antisymmetric structure function $g_1(x, Q^2)$ obtained in reference \cite{Kovensky:2018xxa} to data from SMC \cite{SpinMuon:1998eqa}, E143  \cite{E143:1998hbs}, COMPASS \cite{COMPASS:2010wkz,COMPASS:2015mhb,COMPASS:2017hef} and HERMES \cite{HERMES:2006jyl} collaborations for the proton.

\begin{figure}[h]
\includegraphics[scale=0.9]{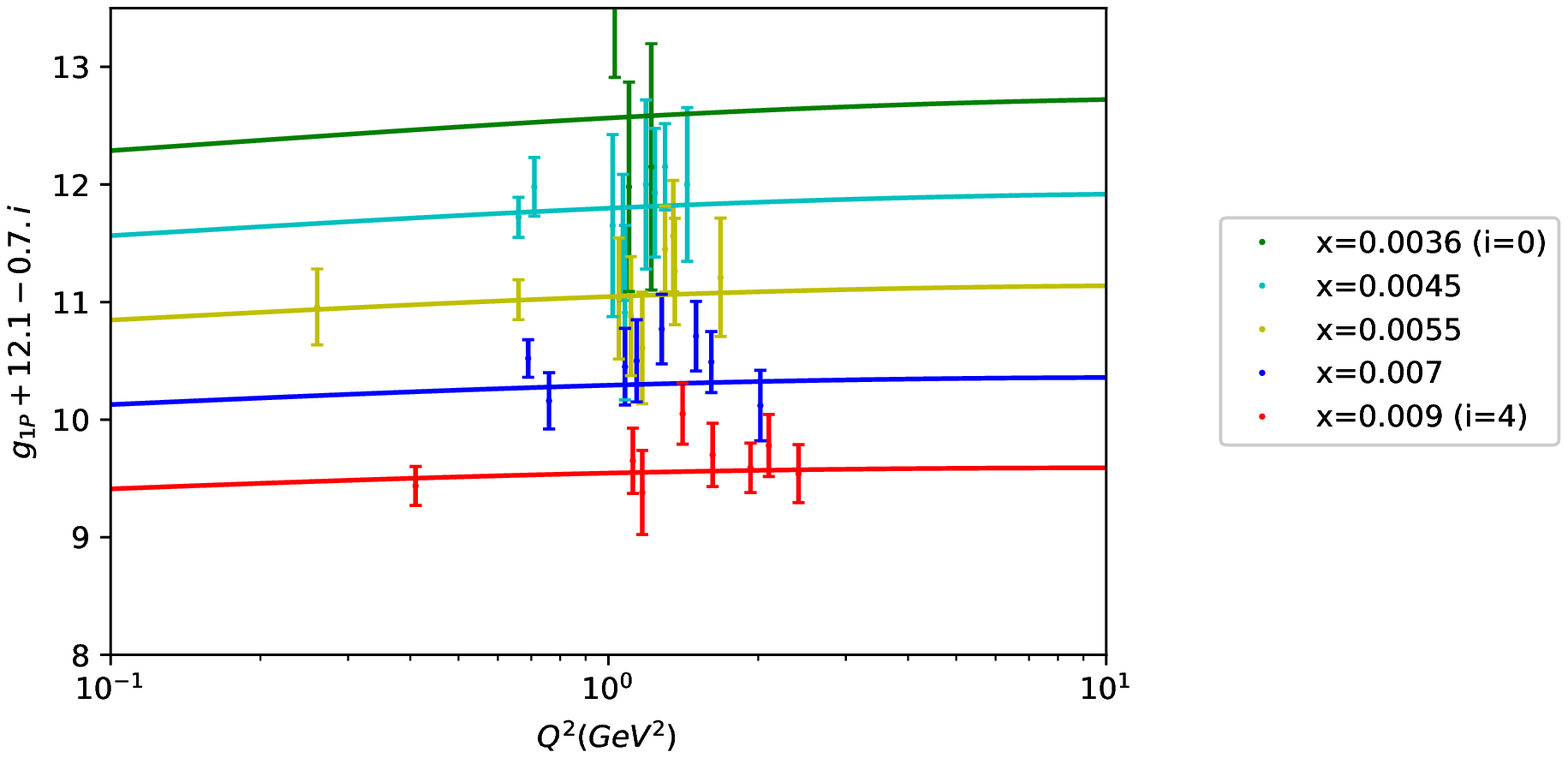}
\caption{\small Best fit of the antisymmetric structure function  $g_1^{\text{BPST}_{\text{HW}}}(x, Q^2)$ from expression (\ref{g1BPSTHW}) in the ranges $0 < x < 0.01$ and 0.1 GeV$^2< Q^2 < 400$ GeV$^2$ to data from SMC \cite{SpinMuon:1998eqa}, E143  \cite{E143:1998hbs}, COMPASS \cite{COMPASS:2010wkz,COMPASS:2015mhb,COMPASS:2017hef} and HERMES \cite{HERMES:2006jyl} collaborations. The parameters $\rho$, $z_0$ and $Q'$ have been obtained from the $F_2^{\text{BPST}_{\text{HW}}}(x, Q^2)$ fit. Thus, there is only one free parameter $C$ to fit to a set of 56 points in total from these collaborations. Notice that for each value of $x$ we add a constant $C_i = 12.1 - 0.7 i$ to the $g_1^P$, which goes from 0 ($x = 0.0036$) to 4 ($x = 0.009$). }
\label{fig-sk-DIS}
\end{figure}

Thus, we fit $g_1^{\text{BPST}_{\text{HW}}}(x, Q^2)$ from expression (\ref{g1BPSTHW}) in the ranges $0 < x < 0.01$ and 0.1 GeV$^2< Q^2 \leq 400$ GeV$^2$, for which there are available data. The function (\ref{g1BPSTHW}) has three parameters $\rho$, $z_0$ and $Q'$, which we have already obtained by fitting to $F_2$ data. Therefore, there is only one free parameter left, $C$, to fit to all the data of $g_1^P$. In total we have considered 56 data to fit $C$ in equation (\ref{g1BPSTHW}). The corresponding results are presented in the first line of table 2. $\chi^2_{\text{d.o.f}}$ is 1.14. Also, we have obtained $\chi^2_{\text{total}}=62$, $N_{d.o.f.}=55$ and the $P$-value is $0.24$. Figure \ref{fig-sk-DIS} shows the results of this fit in comparison with data as explained in the caption.
Next, we consider the sieving procedure neglecting points such that their individual $\Delta\chi^2_i$ is larger or equal to 7. This eliminates only 2 point. The corresponding figure is similar to figure \ref{fig-sk-DIS}. This is indicated in the second line of table 2. It is important to emphasize that while in reference \cite{Kovensky:2018xxa} there have been included only 30 points from the COMPASS collaboration with a restriction to small values of $Q^2$, in our present fits discussed here and displayed in figure \ref{fig-sk-DIS} we consider the whole range of available $Q^2$ for this measurement. This includes 56 experimental points, and the fit presented in line 1 of table 2 is very good. On the other hand, for the second fit in line 2 we have obtained $\chi^2_{\text{d.o.f}}$ is 0.94, which is an indication of over fitting. Also for this case we have obtained $\chi^2_{\text{total}}=50$, $N_{d.o.f.}=53$ and $P=0.59$.

In addition, we have done a fit by considering the four parameters free, i.e. with values of $\rho$, $z_0$ and $Q'$ not to be fixed by fitting to $F_2$. In this case $\chi^2_{\text{d.o.f.}}$ gives 0.6, which is a signal of over-fitting. The idea of this fit was to compare the values of the parameters with those presented in table 2.
\begin{center}
\begin{tabular}{ |c|l|c|c|c|c|c|c|c|c| }
\hline
& {\small Model} & $x$ & $N_p$& sieving & $\rho$ & $C$ & $z_0$ & $Q'$ & $\chi^2_{\text{d.o.f.}}$ \\
&       & range & exp. &  $\Delta\chi^2_{\text{max}}$    &          &      & [GeV$^{-1}$]     &  [GeV]    & \\
\hline \hline

1 & {\small hard-wall}  & $<0.01$ & 56    & No &0.7729  & 0.0145  & 4.894  &0.4715  & 1.14 \\ 

& {\small BPST}    &         &     &    &  & $\pm0.0015$ &  &    &\\
\hline
2 & {\small hard-wall}  & $<0.01$ & 54    & $7$  &0.7729 & 0.162   & 4.894  &0.4715  & 0.94 \\
& {\small BPST}    &         &     &     &  &$\pm0.0014$   &  &     &\\
\hline
3 & {\small hard-wall} & [0.01, 0.1] & 69 & No &0.9194  &0.064  & 3.751 &0.4782  & 2.69 \\  
& {\small BPST}          &         &     &    &  &$\pm0.003$  &  &    &\\
\hline
4 & {\small hard-wall} & [0.01, 0.1] & 60 & $6$ &0.9194 & 0.062  &3.751  &0.4782   & 1.37 \\
& {\small BPST}     &         &     &     &  &$\pm 0.002$ &  &   & \\
\hline
5 & {\small hard-wall} & [0.01, 0.1] & 55 & $4$ &0.9194 & 0.062  &3.751  &0.4782  & 1.15\\
& {\small BPST}     &         &     &     &  &$\pm 0.002$ &  &   & \\
\hline
\end{tabular}
\vspace{0.5cm}
\end{center}
{\small Table 2.  Main results of this work for the fit of $g_1(x, Q^2)$. The first two lines correspond to fits to data within the range $0 < x <0.01$. In lines 3, 4 and 5 we display the corresponding fits in the range $0.01 < x < 0.1$. We consider the range 0.1 GeV$^2 < Q^2 < 400$ GeV$^2$ from SMC \cite{SpinMuon:1998eqa}, E143  \cite{E143:1998hbs}, COMPASS \cite{COMPASS:2010wkz,COMPASS:2015mhb, COMPASS:2017hef} and HERMES \cite{HERMES:2006jyl} collaborations. }

%
%
\subsection{$g_1^P$ at intermediate $x$ values}\label{S-4-2}
%

\begin{figure}[h] 
\centering
\includegraphics[scale=0.9]{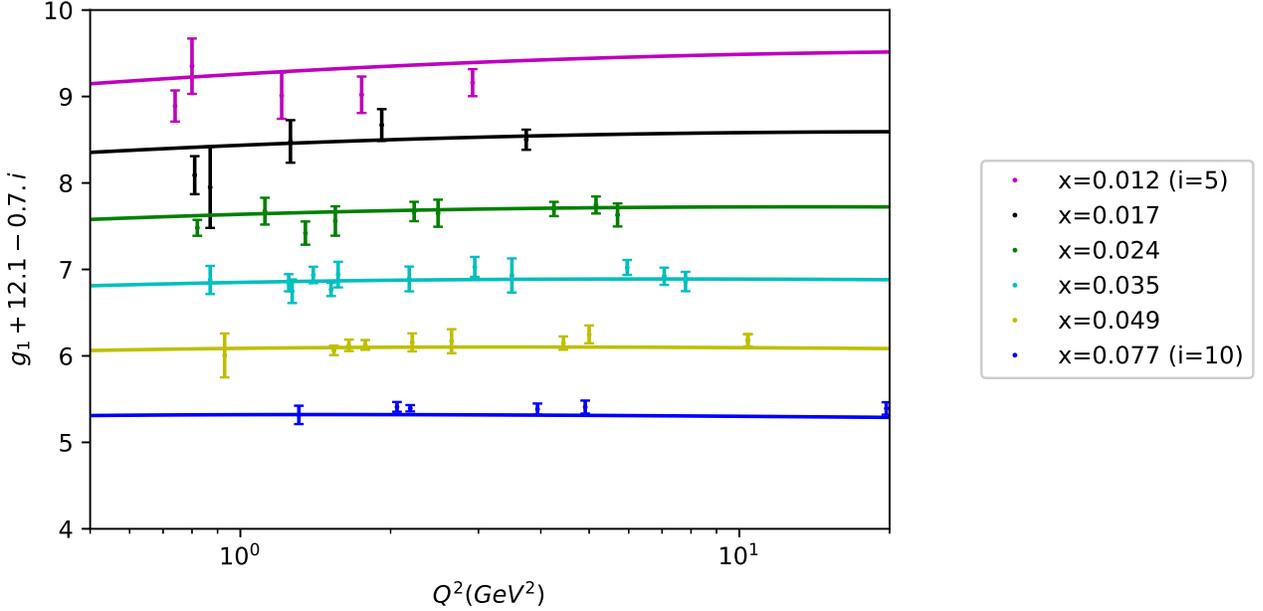}
\caption{{\small Best fit of the antisymmetric structure function  $g_1^{\text{BPST}_{\text{HW}}}(x, Q^2)$ from expression (\ref{g1BPSTHW}) in the ranges $0.01 < x < 0.1$ and 0.1 GeV$^2< Q^2 < 400$ GeV$^2$ to data from SMC \cite{SpinMuon:1998eqa}, E143  \cite{E143:1998hbs}, COMPASS \cite{COMPASS:2010wkz,COMPASS:2015mhb,COMPASS:2017hef} and HERMES \cite{HERMES:2006jyl} collaborations. The parameters $\rho$, $z_0$ and $Q'$ have been fixed from the $F_2^{\text{BPST}_{\text{HW}}}(x, Q^2)$ fit. Thus, as in the previous figure, there is only one free parameter $C$ to fit to 55 points from these collaborations.} }
\label{fig-g1-q-mx}
\end{figure}

We proceed in analogous way as in section \ref{S-3-2-1}, i.e. by considering the exchange of a second single BPST Pomeron in the intermediate range $0.01 < x < 0.1$, and using the same strategy but now considering another single holographic Pomeron exchange related to a Reggeized gauge field in the bulk. We have taken the parameters $\rho$, $z_0$ and $Q'$ from the fit of the BPST Pomeron of section \ref{S-3-2-1}. For this intermediate range of the Bjorken parameter and $0.1 < Q^2 < 400$ GeV$^2$ there are 69 points. All parameters are indicated in the third line of table 2. Since $\chi^2_{\text{d.o.f}}=2.69$ is large (while $\chi^2_{\text{total}}=183$ and $N_p=68$), we implement a sieving, with $\Delta\chi^2_{\text{max}}=6$, which excludes 9 experimental points out of the original 69, and it leads to $\chi^2_{\text{d.o.f.}}=1.37$ (while $\chi^2_{\text{total}}=81$, $N_p=59$ and $P=0.03$). Finally, if we set $\Delta\chi^2_{\text{max}}=4$, there are 55 points while $\chi^2_{\text{d.o.f}}=1.15$. In this way, we can observe that as we reduce the value of $\Delta\chi^2_{\text{max}}$ in the sieving procedure, the value of $\chi^2_{\text{d.o.f}}$ becomes closer to one, but the number of excluded experimental points turns out to be more significant. In this case we obtain $\chi^2_{\text{total}}=62$, $N_p=54$ and $P=0.21$. All parameters are displayed in table 2. Figure \ref{fig-g1-q-mx} shows $g_1$ as a function of $Q^2$ for $0.01 < x < 0.1$. 

%
\subsection{$g_1^P$ at $0.1<x<1$ from type IIB supergravity}\label{S-4-3}
%

As in section \ref{S-3-3} the model based on type IIB supergravity on AdS$_5 \times S^5$ is not able to describe data for $g_1^P$ at $0.1<x<1$. The reason is that we cannot model a dynamical baryon in terms of fundamental quarks in terms of these kind of models. This is because, these type of models do not contain flavour Dp-branes, but, even if we would include flavour Dp-branes, the baryon mass is proportional to $N_c$ in the large-$N_c$ limit. On the other hand, it would be interesting to investigate in the range $0.1<x<1$ what happens with for supergravity dual models in the Veneziano limit \cite{Nunez:2010sf,Jarvinen:2011qe}. The Veneziano limit \cite{Veneziano:1979ec} implies that both the number of colour degrees of freedom $N_c$, as well as the number of flavours $N_f$ are taken very large, but their ratio is kept constant.

%
\section{$A_1^P$ at low $x$ from the holographic Pomeron}\label{S-5}
%

In this section we investigate the virtual Compton scattering asymmetry of the proton $A_1^P$ for low-$x$ values from $g_1$ and $F_2$ discussed in previous sections.

The relation of $A_1^P$ with the structure functions $F_2^P$ and $g_1^P$ is given by the expression
\begin{eqnarray}
A_1^P=2 x \ (1+R) \ \frac{g_1^P}{F_2^P} \, ,
\label{A1p_ec}
\end{eqnarray}
where $R$ is defined from the following relation involving the  longitudinal and transversal cross sections, 
\begin{equation}
    R = \frac{\sigma_L}{\sigma_T} = \frac{F_L}{2xF_1} \, .
\end{equation}
%

We propose to fit $A_1^P$ with a constant $R$ in equation (\ref{A1p_ec}), with $F_2$ and $g_1$ obtained from our previous fits in this work in terms of the holographic Pomeron with a hard-wall. The idea is that, although $R$ is a function of $x$ and $Q^2$, since its variation is smooth \cite{Whitlow:1990dr}, in principle, one may consider to carry out two fits to data of $A_1^P$ in the parametric regions $x < 0.01$ and $0.01 < x < 0.1$, where we use the parameters already obtained from  $F_2$ and $g_1$ in previous sections. Thus, we calculate $R$, which is assumed to be a constant. We consider the following form to calculate the virtual Compton scattering asymmetry
\begin{equation}
A_1^P(x, Q^2)= 2x \ (1+R) \ \frac{g^{\text{BPST}_{\text{HW}}}_{1P}(x, Q^2)}{F_{2P}^{\text{BPST}_{\text{HW}}}(x, Q^2)} \, . 
    \label{A1p_fit_2}
\end{equation}
There are experimental points for $A_1^P$ for the proton in different regimes of $x$ and $Q^2$ from \cite{COMPASS:2010wkz} and \cite{SpinMuon:1998eqa} collaborations.

In the region for $x<0.01$ and $0<Q^2<10$ GeV$^2$ we have 32 experimental points of $A_1^P$. As already commented, the parameters for the structure functions $F_2$ and $g_1$ are taken from our previous fits in this work. In particular, we take the values $g_0^2= 103.73$, $\rho= 0.7729$, $Q'=0.4715 \ {\text {GeV}^2}$ and $z_0 = 4.894$. Thus, by using equation  (\ref{A1p_fit_2}) we obtain: 
\begin{eqnarray}
C= 0.0145 \, , \nonumber
%
\end{eqnarray}
with a very low value $\chi^2_{\text{d.o.f}} = 0.53$, while $R=0.97 \pm 0.22$. This value of $R$ is out of the expected range. For instance, in \cite{E143:1998nvx} it was measured $R=\sigma_L/\sigma_T$ for $0.03 < x < 0.1$ and $1.3 < Q^2 < 2.7$ GeV$^2$. They considered increasing values of both $x$ and $Q^2$, and within these narrow ranges $R$ decreases
from 0.45 to 0.17. In addition, in \cite{Whitlow:1990gk} $R$ has been fitted for $0.1 \leq x \leq 0.9$ and $0.6 \leq Q^2 \leq 20$ GeV$^2$, and it decreases from 0.2 to 0.1. We do not have a clear explanation for this behaviour, considering that in principle it should be smaller for very small values of the Bjorken variable. In this sense, one should take this value of $R$ with caution since this is a poorly known quantity \cite{COMPASS:2017hef}. On the other hand, the very low value of $\chi^2_{\text{d.o.f}}$ indicates that this fit is no good in this region.

On the other hand, for intermediate values $0.01<x<0.1$, with $0<Q^2<10$ GeV$^2$, there are 38 points. Then, we obtain
\begin{equation}
C=0.062   \   .
\end{equation}
having used the following set of parameter from the $F_2$ and $g_1$ fits
\begin{equation}
g_0^2=157.96 \ \ \ \    \rho= 0.9176 \ \ \ \ Q'=0.47 \ {\text {GeV}}^2 \ \ \ \ z_0 = 3.75 \ \ \   .
\end{equation}

\begin{figure}[H]
\centering
\includegraphics[scale=0.9]{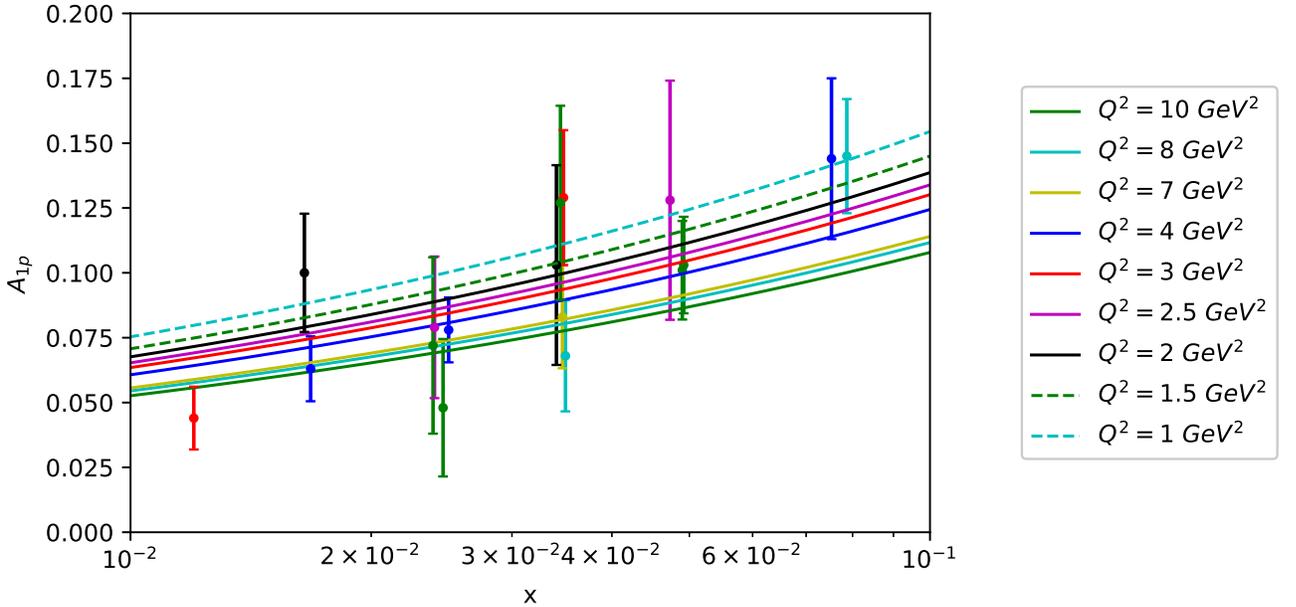}
\caption{{\small Proton virtual Compton scattering asymmetry as a function of the Bjorken parameter for different values of $Q^2$. Experimental data correspond to \cite{COMPASS:2010wkz} and \cite{SpinMuon:1998eqa} collaborations.}} 
\label{A1P_fit}
\end{figure}

By implementing the sieving procedure with $\Delta\chi^2_{\text{max}}=6$, it has been obtained a $\chi^2_{\text{d.o.f}}=1.11$ which is very good, and $R = 0.37 \pm 0.06$. This an expected value for $R$ for these values of $x$. Figure \ref{A1P_fit} shows experimental data of $A_1^P$ as a function of the Bjorken parameter, for different values of $Q^2$ together with the corresponding best fit for intermediate values of $x$. Taking into account the error bars, it is observed a little dependence on $Q^2$, and the curves corresponding to fits for different $Q^2$ values reproduce the trend of experimental data.
These results are compatible with figure 5.6 of reference \cite{Silva:2017}, which shows a dependence of $R$ with $x$ and $Q^2$, related to data from COMPASS collaboration. In this case, for small-$x$ values $R(x)$ seems to develop an $x$-dependence, while for intermediate values of $x$ it seems to be a constant $R \approx 0.4$. Our results for intermediate values of $x$ are compatible with this constant behaviour.

%
\section{Conclusions}\label{S-6}
%

In this work we have presented several results, where the common idea is to investigate how well string theory holographic dual models fit experimental data related to the proton structure functions. Moreover, in the case of polarized DIS, the formulas used are also valid for a domain of very small $x$ and $Q^2>10$ GeV$^2$, where it is expected to have precision measurements by when the Electron-Ion Collider starts its experimental program. Thus, the predictions we discuss in this work will be even more interesting in the forthcoming years, specially for polarized structure functions of the proton.

In the introduction and in section \ref{S-2} we have very briefly reviewed the basic aspects of the formalism needed to understand the holographic dual models we have used to compare with data. Then, in section \ref{S-3} we have extended the fits of $F_2^P(x, Q^2)$ from 249 to around 500 points from different collaborations, obtaining a very good fit, described in table 1 and in that section. The studied range of $x$ is ten times larger than the range considered in previous works. This is very interesting since it tells us that the BPST Pomeron has the ability to describe the parametric values of the Bjorken variable where, in principle (and perhaps naively), one would had expected that string theory scattering amplitudes provide the relevant contribution. Our results show that the BPST Pomeron works well. On the other hand, the string theory scattering amplitudes produce a $(\Lambda^2/Q^2)^{\tau-1}$ factor, which behaves opposite to the trend of experimental data. From the values obtained for the fits of $F_2^P(x, Q^2)$ presented in table 1, we can observe that $\rho=2/\sqrt{\lambda_{\text{'t Hooft}}}$ increases from values of the Bjorken parameter $x < 0.01$ towards the range $[0.01, 0.1]$. This shows a trend for the coupling $\lambda_{\text{'t Hooft}}$ which decreases as $x$ increases. Taking into account that $x \ s \approx Q^2$, at fixed center of mass energy $\sqrt{s}$ the Bjorken parameter becomes proportional to $Q^2$. It therefore means that by increasing $x$ in these conditions, we may expect that the 't Hooft coupling decreases, which is consistent if $Q^2$ increases. The values of the $z_0$ IR cut-off sightly decreases as $x$ increases, meaning that the IR energy cut-off increases less than 2 percent (if we compare lines 2 and 9 of table 1). The variation of $Q'$, which is related to the proton mass, is about 6.5 percent. The larger variation among these parameters is associated with the $g_0^2$ overall constant, which is just a normalization. From this analysis one may infer that the results obtained using two different BPST Pomerons, one for exponentially small $x$ and another for moderately small values of the Bjorken parameter, are consistent with experiments. This suggests that there should be a dependence of the holographic Pomeron with $\lambda_{\text{'t Hooft}}(Q^2)$, not for the superconformal ${\cal {N}}=4$ SYM theory, but for a gauge theory with logarithmic running as in QCD. One should keep in mind that each BPST Pomeron which we have used has only four free parameter to fit more than 200 data for each range of the Bjorken variable.

We have also investigated the situation of polarized DIS. In this case there are two very relevant quantities, namely: $g_1^P(x, Q^2)$ and the virtual Compton scattering asymmetry $A_1^P(x, Q^2)$. For $g_1^P(x, Q^2)$ the number of points we have considered is about twice the number of points considered in previous studies. The results of the fits using a holographic Pomeron for $x < 0.01$ and a second one for the range $[0.01, 0.1]$ are very good. This reinforces the idea commented in the previous paragraph about the need of considering a running coupling in the holographic Pomeron. The experimental data for polarized DIS are in the range 0.1 GeV$^2 < Q^2 <$ 400 GeV$^2$. Moreover, the formulas we have considered for this holographic Pomeron, which are based on the exchange of a single Reggeized gauge field in the AdS$_5 \times S^5$ bulk, are valid for small $x$ values and $Q^2$ corresponding to the polarized DIS at the EIC. Thus, our predictions will be very interesting for polarized data to be measured at the EIC. In this case, there is only one remaining free parameter to fit over 50 experimental data of $g_1(x, Q^2)$.

Also in this context, with the values of the parameters corresponding to the fits of $F_2$ and $g_1$, we have calculated the virtual Compton scattering asymmetry $A_1^P$ and then compared this prediction with experimental data. In this case we have also obtained a very good level of agreement. For this quantity, we have also obtained predictions for future EIC measurements for $Q^2 >$ 10 GeV$^2$ and very small $x$ values.

Very important efforts have been carried out in the last years to understand the origin of the proton spin. In this search the
EIC was conceived to be the most powerful collider to achieve this goal, through measurements with high precision and wide range of $Q^2$. In the case of the antisymmetric structure function of the proton $g_1^P$, it is also expected to obtain extremely precise measurements at values of $x$ down to $10^{-5}$ \cite{Accardi:2012qut,AbdulKhalek:2021gbh}. Particularly, related to the expectations of the EIC program concerning polarized scattering processes, very interesting results have been obtained for $g_1^P$ in reference \cite{Borsa:2020lsz}. They generated pseudo-data calculating that function and other observables. They carried out an extrapolation to the kinematic region of the expected EIC measurements of $g_1^P$. During the extrapolation from measured data for $x$ in $[10^{-2}, 1]$ to the expected new experimental data from EIC, it its assumed that the parton distribution functions have certain form, and also their $Q^2$ dependence is dictated by DGLAP evolution. These two hypotheses, which are instrumental in the analysis developed in reference \cite{DeFlorian:2019xxt}, lead to predictions for extremely low $x$ values, that can be compared with the results of our calculations using the holographic Pomeron.

A different approach has been developed by Kovchegov, Pitonyak and Sievert \cite{Kovchegov:2016weo}, also obtaining a different prediction for $g_1^P$ at the EIC range for low $x$ and the values of $Q^2$. Their evolution equations consider the polarized color-dipole scattering amplitude toward small values of $x$. This procedure leads to predictions for helicity parton distribution functions and for $g_1^P$ at small $x$ from perturbative QCD.

~

At this point, an important question is why the holographic Pomeron works. For very low values of $x$, say smaller than 0.01, we have shown that experimental data of several observables are described well. In order to understand it, we very briefly recall the argument of Brower, Polchinski, Strassler and Tan \cite{Brower:2006ea}. For simplicity, let us assume conformal dynamics. Then, consider the 2 to 2 scattering amplitude of states of a conformal field theory (for this case ${\cal {N}}=4$ SYM theory). This process is dual to the 2 to 2 scattering of closed strings in type IIB string theory in the AdS$_5 \times S^5$ background. Using the metric (\ref{metric1}), the radial coordinate $z$ is related to the energy scale of the ${\cal {N}}=4$ SYM theory at the AdS boundary ($z \rightarrow 0$ which corresponds to the UV limit of the field theory). It has been shown in \cite{Brower:2006ea} that the BPST Pomeron kernel for the Mandelstam variable $t=0$, is given by the following expression
\begin{equation}
{\text {Kernel}}_{\text {BPST}}(z, z', s) = \frac{s^{j_0^{\text {BPST}}}}{(4 \pi D_{\text {BPST}} \ \ln s)^{1/2}} \ \exp[-(\ln z'- \ln z)^2/(4D_{\text {BPST}} \ \ln s)] \, , \label{BPST-kernel}
\end{equation}
where there is a diffusion constant: $D_{\text {BPST}}=1/(2 \lambda_{\text {'t Hooft}}^{1/2}) + {\cal {O}}(\lambda_{\text {'t Hooft}}^{-1})$ and $j_0^{\text {BPST}}=2-2/\lambda_{\text {'t Hooft}}^{1/2}+ {\cal {O}}(\lambda_{\text {'t Hooft}}^{-1})$. Now, let us compare this expression with the BFKL Pomeron kernel, which obviously has a very different interpretation, being obtained from perturbative calculations in the large $N_c$ limit. The corresponding single BFKL-Pomeron exchange scattering amplitude between two hadrons whose structure is described by the impact factors $\Phi_1(p_\bot)$ and $\Phi_2(p'_\bot)$, respectively, is given by
\begin{equation}
\int \frac{dp_\bot}{p_\bot} \ \int \frac{dp'_\bot}{p'_\bot} \
\Phi_1(p_\bot) \ {\text {Kernel}}_{\text {BFKL}}(p_\bot, p'_\bot, s) \ \Phi_2(p'_\bot) \, ,
\end{equation}
where $p_\bot$ is the transverse momentum with which the first hadron is probed by the BFKL Pomeron, and similarly $p'_\bot$ is the transverse momentum with which the second hadron interacts with it. A good approximation for this kernel gives 
\begin{equation}
{\text {Kernel}}_{\text {BFKL}}(p_\bot, p'_\bot, s) \approx \frac{s^{j_0^{\text {BFKL}}}}{(4 \pi D_{\text {BFKL}} \ \ln s)^{1/2}} \ \exp[-(\ln p'_\bot- \ln p_\bot)^2/(4D_{\text {BFKL}} \ \ln s)] \, , \label{BFKL-kernel}
\end{equation}
where now the diffusion constant becomes
\begin{equation}
D_{\text {BFKL}}=\frac{7 \zeta(3)}{8 \pi^2} \lambda_{\text {'t Hooft}} \, ,
\end{equation}
and $j_0^{\text {BFKL}}=1+ \frac{\ln 2}{\pi^2} \lambda_{\text {'t Hooft}}$. It is effectively a diffusion kernel, with diffusion in $\ln p_\bot$, over a diffusion time $\tau \approx \ln s$. It is important to stress that the BFKL Pomeron is valid for small $\lambda_{\text {'t Hooft}}$, while the BPST Pomeron was derived for large coupling. Now, let us compare the BPST Pomeron kernel (\ref{BPST-kernel}) with the BFKL Pomeron one (\ref{BFKL-kernel}). One can see the identification of $1/z$ with $p_\bot$. Notice that in both cases there is the same diffusion time. Thus, there is a connexion between the Reggeized gluon of pertubative QCD and the Regge limit for a propagation of a closed string in AdS$_5 \times S^5$ \cite{Brower:2006ea}.

Recall that a single holographic Pomeron exchange dominates at large $N_c$ and large center of mass energy $\sqrt{s}$. At finite $N_c$ the multi-Pomeron exchange could become the leading contribution as $s$ increases. In this work, we have considered only a single holographic Pomeron exchange, and the results of the comparison with experimental data are very good. In any case, the question of the role of $1/N_c^2$ contributions is very important. In the context of the BPST Pomeron, it was studied by considering the eikonal approximation in \cite{Cornalba:2006xm,Cornalba:2007zb,Brower:2007qh,Brower:2007xg,Cornalba:2008qf}. For $F_2$ it was investigated in \cite{Brower:2006ea}, concluding that the onset of saturation occurs for very small values of $Q^2$.

Another very interesting result is the fact, already commented, that the expressions  we use for the holographic Pomeron are valid for extremely small-$x$ values, and for $Q^2$ typically larger than 1 GeV$^2$, which avoids saturation effects, and smaller than the domain where electroweak interactions become relevant. In this sense, our predictions offer the possibility to investigate and compare with the situation of polarized DIS at the EIC.

For intermediate values of the Bjorken parameter it was derived the DIS hadronic tensor using a local approximation for the string theory scattering amplitudes for glueballs in \cite{Polchinski:2002jw}, for spin-1/2 hadrons in \cite{Kovensky:2018xxa}, and for mesons in \cite{Koile:2014vca}.
First of all, let us emphasize that one crucial difference between the holographic dual description of mesons is that simply using probe flavour Dp-branes a dynamical meson can be described in terms of fundamental open strings attached to the flavour Dp-brane. On the other hand, glueballs are described as closed strings. However, there are no dynamical baryons constructed from fundamental open strings. This is because they must end on a baryon vertex, for instance a D5-brane wrapping the $S^5$ \cite{Witten:1998xy} (also \cite{Gubser:1998fp,Polchinski:2000uf}). Thus, their masses scale as $N_c$, which in the gauge/string theory approximation is very large. In this discussion, we neglect corrections of the order $1/N_c^2 \equiv g_{\text{string}}^2$, related to the string coupling, corresponding to the world-sheet topology of the torus.  The issue is that in the string theory scattering amplitude calculation a certain approximation is made, and consequently, it turns out that diffusion in the radial coordinate is neglected. The result for the spin-1/2 fermions is that the $(\Lambda^2/Q^2)^{\tau-1}$ power behaviour is the opposite to the trend indicated by experiments. On the other hand, it turns out that we can fit data very well by considering a single-holographic Pomeron exchange. As it was explained in this section, this is compatible with a running coupling dependence of the holographic-Pomeron kernel.

The supergravity dual description related to the $s$-channel of forward Compton scattering cannot be used to compare with data. The reason is that as explained in the previous paragraph, there are not top-down holographic models, even including flavours, with the ability of representing dynamical baryons composed of fermions in the fundamental representation in the planar limit. Thus, since for $0.1 < x < 1$ DIS is dominated by valence quarks, apparently there is no known way top-down dual supergravity models give a correct description of DIS.

We should mention that in terms of the so-called holographic QCD (AdS/QCD or bottom-up models) many important hadronic properties have been investigated. Related to hadron structure it is interesting to mention, for instance, the calculations with the soft Pomeron in  holographic QCD developed in reference \cite{Ballon-Bayona:2015wra}. A very interesting result derived from holographic QCD is presented in \cite{Ballon-Bayona:2017vlm} where the authors fit $F_2$ as a function of the Bjorken parameter for 249 points from HERA, for $x < 0.01$ and $0.1 < Q^2 < 400$ GeV$^2$ with a $\chi^2_{\text{d.o.f}}=1.7$, which better than the corresponding one for the BPST Pomeron in this region (see line 1 of our table 1). While in the BPST Pomeron fits (lines 1-4 of table 1) there are 4 free parameter, in the case of reference \cite{Ballon-Bayona:2017vlm} there are 9 free parameters. In a previous paper \cite{Amorim:2018yod}, it was considered holographic QCD with non-minimal coupling contributions, obtaining $\chi^2_{\text{d.o.f}}=1.1$ using 14 parameter. Also, the BPST Pomeron in holographic QCD has been studied in  \cite{Watanabe:2012uc} for the calculation of $F_2$. Generalized parton distribution functions of quarks and gluons in holographic QCD have been studied for different observables and compared to experiments \cite{Mamo:2022jhp}. With respect to the Veneziano limit that we commented before for top-down models, in \cite{Amorim:2021gat} the authors initiate a study of Regge theory in a bottom-up holographic model for QCD.

~

~

%
\centerline{\large{\bf Acknowledgments}}
%

~

We have been benefited from comments by Mar\'{\i}a Teresa Dova, Jos\'e Goity, Carlos N\'u\~nez and Rodolfo Sassot. We specially thank Rodolfo Sassot for discussions, suggestions and for calling our attention on several references on polarized DIS and the Electron-Ion Collider. This work has been supported in part by the Consejo Nacional de Investigaciones Cient\'{\i}fi\-cas y T\'ecnicas of Argentina (CONICET), the Agencia Nacional para la Promoci\'on de la Ciencia y la Tecnolog\'{\i}a of Argentina (ANPCyT-FONCyT) Grant PICT-2017-1647, the UNLP Grant PID-X791, and the CONICET Grants PIP-UE B\'usqueda de nueva f\'{\i}sica and PICT-E 2018-0300 (BCIE).

\newpage

\appendix

\section{Appendix: Brief comments on the sieving method} \label{Appendix-A}

In the fits to experimental data described in this work we have used the sieving method described by Block in reference \cite{Block:2005qm}. This method contains several steps that we briefly explain below.

%

{\it Step 1 :} In order to carry out a robust fit to all data, one firstly minimizes the Lorentzian squared function 
$\Lambda^2_0$, which is defined as 
\begin{equation}
\Lambda^2_0 = \sum_{i=1}^N \ln{(1+0.18 \ \Delta \chi_i^2(x_i,\alpha))} \, ,
\end{equation}
where $\alpha = \{ \alpha_1 , . . . , \alpha_M \}$ represents the space of parameters corresponding to the function which we want to compare to data. A set of experimental data indicated by  $y=\{ y_1 , . . . , y_N \}$ corresponds to the values of $x=\{ x_1 , . . . , x_N \}$, respectively. In the cases we present in this work $x_i$ can be either the value of the Bjorken parameter or the value of the squared of the virtual-photon momentum transfer, $Q_i^2$, at which the structure function is measured. For instance, in the first situation we have $y_i=F_2(x_i, Q^2)$, which corresponds to an experimental set of data for a certain fixed value of $Q^2$.  

Then, one defines
\begin{equation}
\Delta \chi_i^2(x_i,\alpha)=\left(\frac{y_i-y(x_i,\alpha) }{\sigma_i}\right)^2 \, , \label{DeltaChi2}
\end{equation}
where $y(x_i,\alpha)$ is the theoretical value and $\sigma_i$ is the experimental uncertainty corresponding to that point, $y_i$. This object will allow to quantify how far a certain experimental point $y_i$ lies from the signal, and therefore to identify whether this point can be considered as an ``outlier".
Specifically, if $\chi^2_{min}=\sum_{i=1}^N \ \Delta \chi_i^2(x_i, \alpha)$ obtained after minimizing $\Lambda^2_0$ is satisfactory, then one carries out a conventional $ \chi^2$ fit, and the uncertainties of the parameters $\alpha$ can be calculated. As a standard convention, when $\chi^2_{min}$ is close to the number of degrees of freedom ($N$-$M$), it is considered to be satisfactory. On the other hand, if $\chi^2_{min}$ turns out not to be satisfactory one should follow step 2. 
    
~
       
{\it Step 2:} Then, one uses the {\it regressors} obtained from the fit with $\Lambda_0^2$ and calculate $\Delta \chi_i^2(x_i,\alpha)$ for each one of the $N$ points.

~

{\it Step 3:} Next, a certain value $\Delta \chi^2_{max}$ is set to carry out the sieving. It means that points such that $\Delta \chi_i^2(x_i,\alpha) \geq \Delta \chi^2_{max}$ are to be excluded. The idea is to set $\Delta \chi^2_{max}$ as large a possible in order to eliminate only ``outliers", thus trying to minimize the number of points of the signal which would be excluded arbitrarily by the sieving.

~

{\it Step 4:} With the not-excluded points in the previous step  one calculates the usual $\chi^2_{min}$. This result has to be corrected due to the truncation (elimination) of the points through the discussed sieving mechanism. This correction is implemented in terms of a factor ${\cal{R}}$ which in turn depends upon $\Delta \chi^2_{max}$. If the corrected $\chi^2_{min}$ is aceptable in the conventional sense, then one follows step 5. Otherwise, if the corrected $\chi^2_{min}$ is not aceptable, and if $\Delta \chi^2_{max}$ is not too small, then one can still choose a smaller value of $\Delta \chi^2_{max}$ and return to step 3. The limiting value for the algorithm to work is $\Delta \chi^2_{max}>2$ \cite{Block:2005qm}.

~

{\it Step 5:} From the fit developed in the previous step, using the sieved data one obtains the set of parameters $\alpha$. Then, the uncertainties of such parameters can be calculated from the covariance matrix, which is a $M \times M$ square matrix. Also, this matrix must be corrected multiplying it by a factor $r_{\chi^2}$, depending on $\Delta \chi^2_{max}$. 

~

It is worth to say that the value of $\chi^2_{\text{d.o.f}}$ obtained from the sieving is not necessarily the closer to 1. This is due to the normalization constant, which increases as $\Delta \chi^2_{max}$ decreases. 

~

\begin{figure}[H]
        \centering
        \begin{subfigure}[b]{0.45\textwidth}
            \centering
            \includegraphics[width=\textwidth]{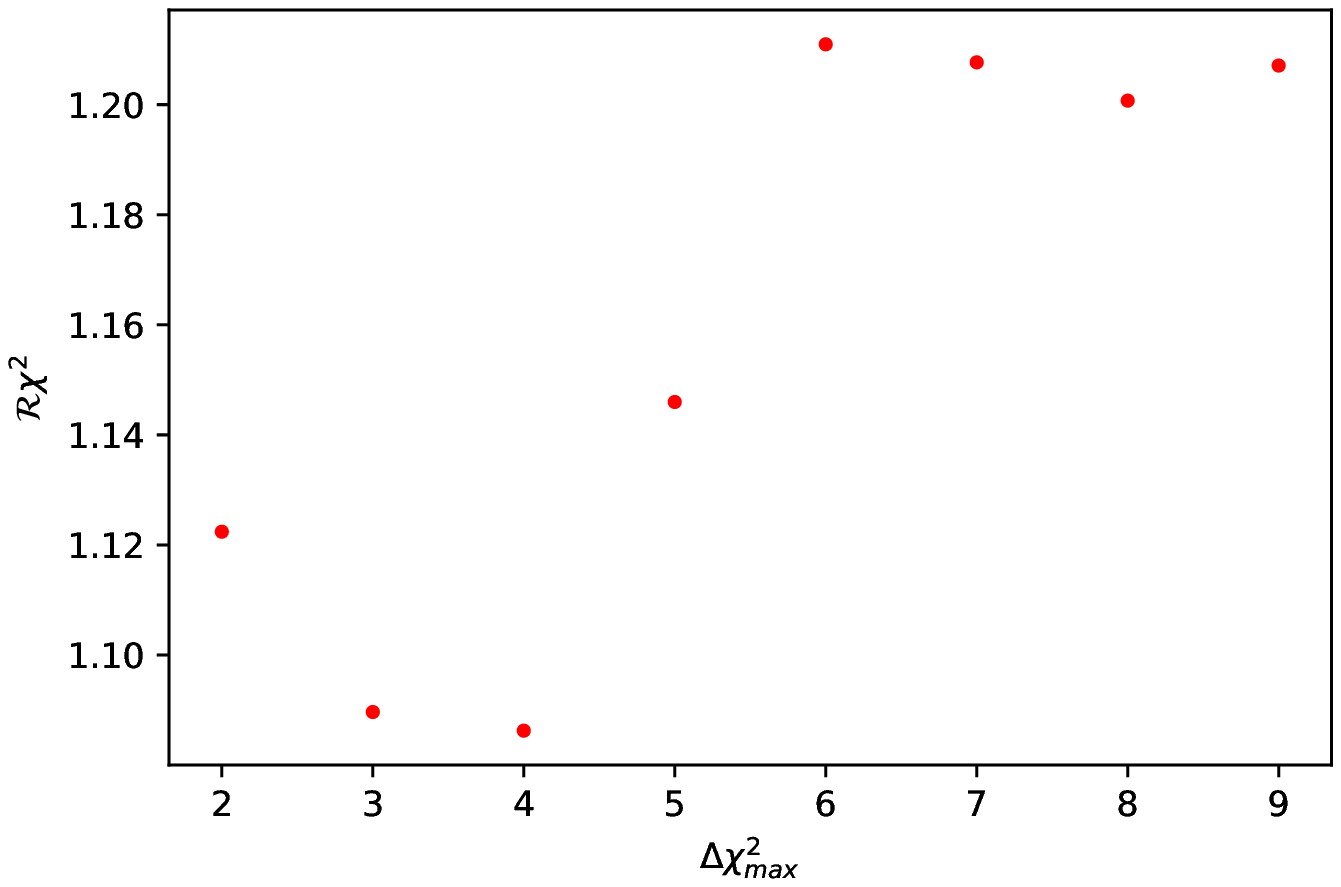}
            \caption[Network2]%
            {{\small}}    
        \end{subfigure}
        \hfill
        \begin{subfigure}[b]{0.45\textwidth} 
            \centering 
            \includegraphics[width=\textwidth]{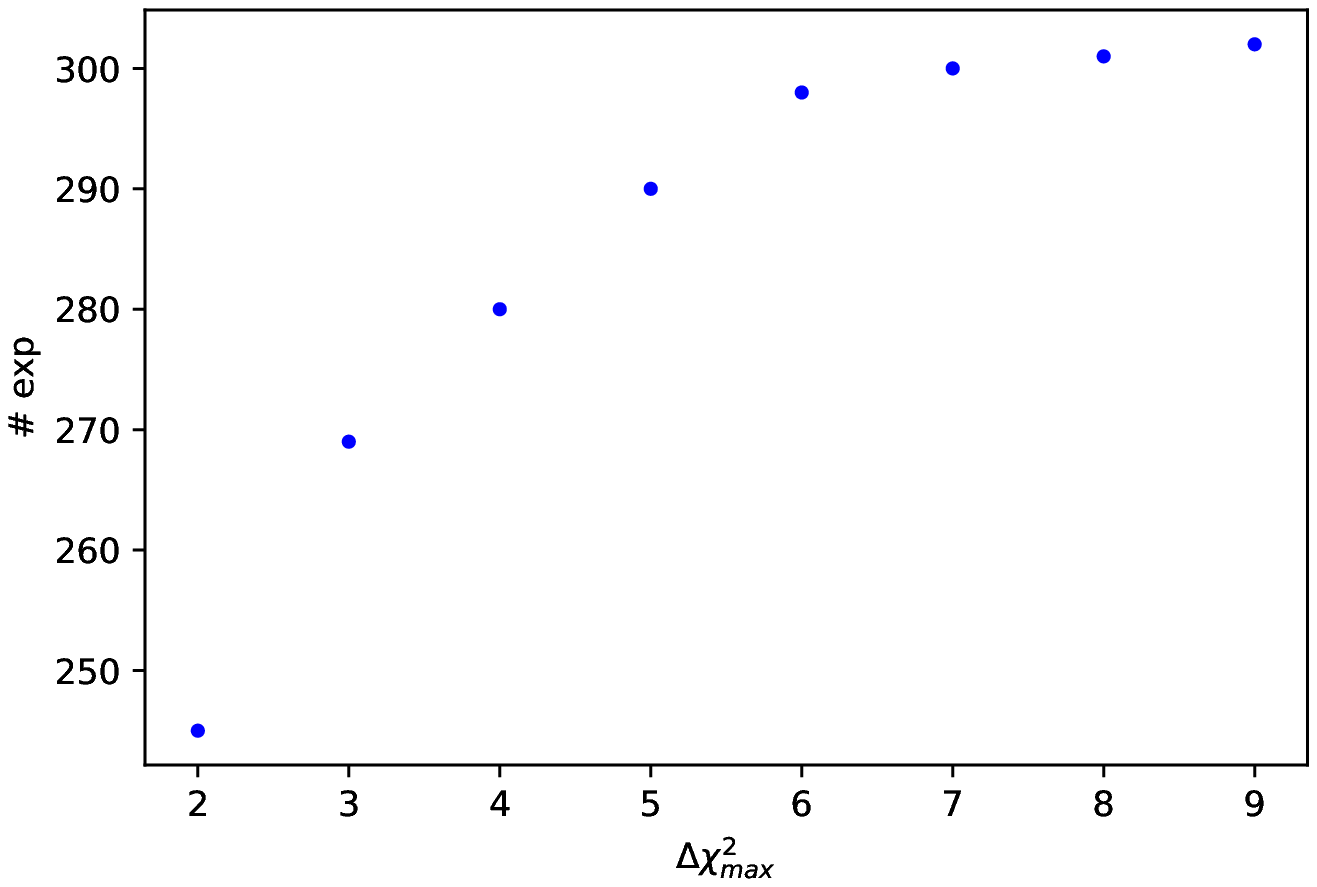}
            \caption[]%
            {{\small}}    
        \end{subfigure}
            \hfill
             \caption[]%
            {{\small} Panel (a) corresponds to the normalized $\chi^2_{\text{d.o.f}}$ as a function of $\Delta \chi^2_{max}$. Panel (b) shows the number of experimental points after sieving as a function of  $\Delta\chi^2_{max}$.} 
            \label{fig_sieving}
    \end{figure}

As an example, figure \ref{fig_sieving} displays the normalized $\chi^2_{\text{d.o.f}}$ (panel (a)) and the number of experimental points (panel (b)), after sieving, as function of $\Delta \chi^2_{max}$ for the fit of $F_2$ in the case of the BPST Pomeron for  $0<x<0.01$. 
We can observe that $\Delta \chi^2_{max}=4$ is a good value for sieving since the number of excluded experimental points is small and $\chi^2_{\text{d.o.f}}$ is close to 1, even closer than for any other $\Delta \chi^2_{max}$. Notice that still for smaller values of $\Delta \chi^2_{max}$ as 2 or 3, which implies excluding more experimental points, the fit does not improve in comparison with the mentioned case for $\Delta \chi^2_{max}=4$.

\end{document}